\documentclass[]{aa}  \usepackage{txfonts}
\usepackage{natbib} \usepackage{graphicx} \usepackage{color}
\newcommand{\Lower}[1]{\smash{\lower 1.5ex \hbox{#1}}}
\newcommand\T{\rule{0pt}{2.6ex}}
\newcommand\B{\rule[-1.2ex]{0pt}{0pt}}

\begin{document}

\title{
Cosmic rays and the magnetic field in the nearby starburst\\
  galaxy NGC\,253 \\ {\Large III. Helical magnetic fields in the nuclear outflow}}

\titlerunning{Helical magnetic fields in the nuclear outflow of NGC\,253}
 
\author{V. Heesen\inst{1} \and R. Beck\inst{2}
\and M. Krause\inst{2} \and R.-J. Dettmar\inst{3}}

\institute{Centre for Astrophysics Research, University of
  Hertfordshire, Hatfield AL10 9AB, UK \\ \email{v.heesen@herts.ac.uk} \and Max-Planck-Institut f\"ur
  Radioastronomie, Auf dem H\"ugel 69, 53121 Bonn , Germany \and
  Astronomisches Institut der Ruhr-Universit\"at Bochum,
  Universit\"atsstr. 150, 44780 Bochum, Germany}

\date{Received 4 July 2011 / Accepted 30 August 2011}
\abstract
{Magnetic fields are good tracers of gas compression by shock waves in the
  interstellar medium.  These can be caused by the interaction of
  star-formation driven outflows from individual star formation sites as
  described in the \emph{chimney} model. Integration along the
  line-of-sight and cosmic-ray diffusion may hamper detection of
  compressed magnetic fields in many cases.}
{We study the magnetic field structure in the central part of the
  nuclear starburst galaxy NGC\,253 with spatial resolutions between $40$
  and $150\,\rm pc$ to detect any filamentary emission associated with the nuclear
  outflow. As the nuclear region is much brighter than the rest of the disc we
  can distinguish this emission from that of the disc.}
{We used radio polarimetric observations with the VLA. New observations at
  $\lambda$3\,cm with $7\farcs 5$ resolution were combined with archive data
  at $\lambda\lambda$ 20 and 6\,cm. We created a map of the rotation measure
  distribution between $\lambda\lambda$ 6 and 3\,cm and compared it with a
  synthetic polarization map.}
{We find filamentary radio continuum emission in a geometrical distribution, which we interpret as the boundary of the NW nuclear outflow cone
  seen in projection. The scaleheight of the continuum emission is $150\pm
  20\,\rm pc$, regardless of the observing frequency.  The equipartition
  magnetic field strength is $46\pm 10\,\mu\rm G$ for the total field and
  $21\pm 5\,\mu\rm G$ for the regular field in the filaments.  We find that
  the ordered magnetic field is aligned along the filaments, in agreement with
  amplification due to compression. The perpendicular diffusion coefficient
  across the filaments is $\kappa_{\rm \perp} = 1.5\times 10^{28}\,{\rm cm^2\,
    s^{-1}} \cdot E({\rm GeV})^{0.5\pm0.7}$.  In the SE part of the nuclear
  outflow cone the magnetic field is pointing away from the disc in form of a
  helix, with an azimuthal component increasing up to at least 1200\,pc
  height, where it is about equal to the total component.  The ordered
  magnetic field in the disc is \emph{anisotropic} within a radius of
  2.2\,kpc.  At larger radii, the large-scale field is regular and of even
  parity.}
{The magnetic filaments indicate an interaction of the nuclear outflow with
  the interstellar medium. The magnetic field is able to collimate the
  outflow, which can explain the observed small opening angle of $\approx
  26\degr$.  Owing to the conservation of angular momentum by the plasma in
  the nuclear outflow, the field lines are frozen into the plasma, and they
  wind up into a helix. Strong adiabatic losses of the cosmic-ray electrons in
  the accelerated outflow can partly explain why the radio luminosity of the
  nucleus lies below the radio-FIR correlation.}

\maketitle
\setcounter{equation}{0} \setcounter{figure}{0}
\setcounter{table}{0} \setcounter{footnote}{0}
\setcounter{section}{0} \setcounter{subsection}{0}
\section{Introduction}
\label{sec:mf_introduction}
Magnetic fields are an important ingredient of the interstellar medium. Not
only do they facilitate star formation, but they also make an important
contribution to the energy budget. The global magnetic field structure of
galaxies can be described by a spiral field in the disc and a poloidal field
in the halo, which takes an X-shape when seen edge-on \citep{krause_08a,
  beck_09a, braun_10a}. Such X-shaped structures may be formed by the
hydrodynamic Al outflows driven by star formation in the disc
\citep{dallavecchia_08a}. On the other hand, a concurring theory is that these
magnetic field structures are part of a large-scale field, possibly generated
by a galactic dynamo \citep{gissinger_09a,hanasz_09a}. Therefore, it is
desirable to understand how the dynamics of the interstellar medium influence
the structure of the magnetic field that is frozen into the gas. The magnetic
field may only be a tracer for interactions but can also influence the
dynamics if its energy density is high enough.

\begin{table*}[tbhp]
\caption{VLA observations presented in this paper.}
\begin{center}
\begin{tabular}{rrrrrrrrrrr}
  \hline\hline\rule{0in}{2ex}
  $\lambda$ & $\nu$ & Array & Resolution & TP noise & PI noise & Dynamic &Date
  &Code & References\\
  $[\rm cm]$ & [GHz] &  &  & \multicolumn{2}{c}{$[\rm \mu Jy\, beam^{-1}]$} &range&&\\\hline\rule{0in}{2ex}
  $20$ & 1.49 & A & $1\farcs 3\times 2\farcs 2$ & $50$ & $-$ & 3600 & Jul.\
  1987 & AU30 & \citet{ulvestad_97a}\\
  $20$ & 1.56 & B$+$C & $7\farcs 5$ & $400$ & $-$ & 2500 & Sep.\ 1990 & AC278
  & \citet{carilli_92a} \\
  $6^*$ & 4.71 &C & $7\farcs 5$ & $100$ & $40$ & 8300 & Jan.\ 1991 & AC278 &
  this paper\\
  $3^*$ & 8.46 & DnC & $7\farcs 5$ & $50$ & $16$ & 13000 & Oct.\ 2009 & AH899
  & this paper\\\hline
\end{tabular}
\end{center}
\small{$^*$The observations at $\lambda\lambda$ 6 and 3\,cm were
merged with Effelsberg single-dish data.} \label{tab:VLA}
\end{table*}
Observations of nearby galaxies show that the large-scale magnetic field is a
very sensitive tracer for interaction in the interstellar medium that is not
seen at any other wavelength. An example for this are the colliding antenna
galaxies that show a maximum of polarized intensity outside of their optical
extents \citep{chyzy_04a}.  Several galaxies in the Virgo cluster show signs
of interaction possibly due to ram pressure compression as sharp edges in the
total power emission and enhanced polarization
\citep{vollmer_07a,vollmer_10a,wezgowiec_07a}.  MHD simulations by
\citet{avillez_00a} of the interstellar medium reveal a highly filamentary
disc-halo interface with transient vertical filaments in the gas and magnetic
fields. The filaments could be regarded as the walls of so-called
\emph{chimneys} through which hot gas can break out from the disc into the
halo when a sufficient number of supernovae occur in a OB association
\citep{norman_89a, maclow_99a}. Farther up in the halo the gas flow may be
driven by the cosmic-ray pressure, which has a larger scaleheight and is not
so much affected by radiation losses
\citep{ipavich_75a,breitschwerdt_91a,breitschwerdt_08a}. Observations of this
phenomenon will be left to the upcoming radio telescopes that can observe at
the very lowest frequencies and are sensitive to the aged cosmic-ray
electrons.  Magnetic fields may, however, already be important for outflows in
the disc.
Searches for small-scale magnetic field structures in galaxies that
trace the filamentary disc-halo interface are rare. Normally the radio
continuum emission is smooth and can be described as two-component exponential
haloes with a thin (300\,pc) and a thick radio (1.8\,kpc) disc
\citep{dumke_98a}. Vertical filaments are observed in M82 with prominent gaps
in the continuum emission, possibly created by vertical magnetic fields, which
prevent horizontal diffusion \citep{reuter_92a,reuter_94a}. But it was not
possible to relate these minima to a particular outflow cone
\citep{wills_99a}. In NGC\,5775 vertical filaments in the radio spectral index
suggest the outflow of young cosmic-ray electrons, but they do not coincide
with the observed radio spurs with vertical magnetic fields
\citep{duric_98a,tuellmann_00a}. In NGC\,1569 a huge H$\alpha$ bubble can be
identified as the edge of a radio spur \citep{kepley_10a}. In NGC\,253 the
halo magnetic field seems to line up tangentially with the edge of the
superbubble filled with soft X-ray emission \citep[][hereafter
 Paper~II]{heesen_09b}.
 NGC\,253 is an archetypical starburst galaxy that has both a superwind and
 a disc wind \citep{heckman_00a, strickland_02a, heesen_09a}. The superwind is
 traced by the hot X-ray emission seen in the halo, indicating the outflow of
 hot gas, possibly supernova-heated. The hot gas is seen in the centre as a
 plume of 600\,pc length emerging to the southeastern (SE) side
 \citep{strickland_00a,bauer_07a}. The northwestern (NW) side is on the
 far side of the disc and thus heavily absorbed in X-ray emission
 \citep{pietsch_00a}. \citet{bauer_07a} detected with sensitive XMM--Newton
 observations the X-ray emission of the actual fluid in the outflow, whereas
 \citet{strickland_00a} used the superior resolution of the \emph{Chandra}
 telescope to resolve the filaments consisting of cooled material. The
 speed of the hot gas could be measured from H$\alpha$ emission lines to be
 $390\,\rm km\,s^{-1}$ by \citet{schulz_92a}. Recent, optical integral-field
 unit, spectroscopy observations could measure the increase in the outflow
 velocity with increasing distance from the nucleus \citep{westmoquette_11a}.
The disc wind that is traced by a radio continuum halo is different from the
nuclear outflow. In \citet[][hereafter Paper~I]{heesen_09a}, we have shown that
the cosmic-ray electrons are transported with a bulk velocity of about
$300\rm\, km\,s^{-1}$ in a mainly vertical direction over the full extent of
the disc. The outflow speed varies only slightly as a function of the
galactocentric radius, further supporting the idea of a disc wind, because it
seems unrelated to the nuclear outflow. A high-resolution study of the
continuum emission by \citet{zirakashvili_06a} of the inner disc showed a very
similar outflow speed.
In Paper~II we investigated the magnetic field structure and found that
it is an axisymmetric spiral in the disc and X-shaped in the halo, as in
several other edge-on galaxies \citep{krause_08a}.  The origin of the X-shapes
is still not clear, but in NGC\,253 it may be related to the interaction of the
disc wind with the superwind.  Simulations of a superwind by
\citet{suchkov_94a} show the formation of huge bubbles. The outer lobes of the
bubbles show an X-shaped structure in projection. If the large-scale magnetic
field is compressed in these lobes and aligned with them, they would also form
an X-shaped structure.
So far, no high-resolution magnetic field maps have been available for
NGC\,253 that would allow us to trace the structure of the X-shaped halo field
back into the inner disc and study its interaction with the nuclear
outflow. The magnetic field maps shown in Paper~II have a resolution of
$30\arcsec$. In this paper, we present new observations with less than
$10\arcsec$ resolution to study the magnetic field structure in the inner disc
of NGC\,253. In Sect.\,\ref{sec:mf_observations_and_data_reduction} we present
our observations and data reduction. In
Sect.\,\ref{sec:structure_of_the_radio_continuum_emission} we discuss the
structure of the radio continuum
emission. Section~\ref{sec:magnetic_field_structure} contains the analysis of
the magnetic field structure measured with
polarimetry. Section~\ref{sec:discussion} contains the discussion and
Sect.\,\ref{sec:conclusions} our conclusions. In this paper we use the
distance from \citet{karachentsev_03a} of $3.94\,\rm Mpc$ ($1\arcsec
=19.1\,\rm pc$), a position angle of $52\degr$, and an inclination angle of
$78\degr$ \citep{pence_81a}.
\section{Observations and data reduction}
\label{sec:mf_observations_and_data_reduction}
\subsection{New VLA observations}
We used radio continuum polarimetric observations with the Very Large Array
(VLA) at $\lambda$3\,cm with a bandwidth of 2 $\times$ 50\,MHz\footnote{The
  National Radio Astronomy Observatory (NRAO) is a facility of the National
  Science Foundation operated under cooperative agreement by Associated
  Universities, Inc.}. The observations were carried out in October 2009 on
three days amounting to 11\,h of total observing time. We used a hybrid
DnC-configuration to get a circular beam at the southern declination of
NGC\,253 ($\delta=-25\degr$). We selected three pointings with about 3.5\,h of
on-source integration time each. The data were calibrated with J0137+331
(3C48) using the \citet{baars_77a} flux scale, and J0025-260 was used as a
secondary (phase) calibrator.
For the data reduction we used the Astronomical Image Processing System
(AIPS)\footnote{AIPS is free software from NRAO.}. For total power (Stokes
$I$), the dynamic range has to be large because of the 1\,Jy nuclear
point-like source. We used self-calibration first with only phase solutions
($2\times$) and subsequently with a solution in phase \& amplitude
($2\times$), which improved our dynamic range. Of the 24 antennas in the
array, 21 were upgraded EVLA antennas and three were VLA antennas. Our final
map improved significantly when we left out the three VLA antennas, which we
did not investigate any further. The r.m.s.\ noise level of $50\,\mu\rm
Jy\,beam^{-1}$ is a factor of 3 higher than the theoretically expected
value. We explain this with the high dynamic range, where in our map the
peak-to-noise ratio is 13000. Since the sensitivity is enough for our analysis
and the polarization is not affected by it, as shown below, we did not try to
improve our map any further.
The polarization leakage (D-terms) was calibrated with {\tt PCAL} using the
secondary calibrator and observing the variation in the polarization as a
function of the parallactic angle. The source intrinsic polarization varies
with the parallactic angle, whereas the instrumental polarization stays
constant, allowing us to separate them from each other. The polarization angle
is calibrated with the known angle of 3C48. The calibrated $(u,v)$-data were
inverted and cleaned until we reached a level of $2\times$ the r.m.s.\ noise
level. We used a robust weighting \citep{briggs_95a} with robust=0, which gave
us a beam size of $7\farcs 5$ both in total power and in
polarization.\footnote{All resolutions are referred to as the half power beam
  width (HPBW).} The three pointings were combined with {\tt LTESS} (part of
AIPS) where a linear superposition is used with a correction for the
attenuation of the primary beam \citep{braun_88a}. We took the part of the
maps into account that was above 7\% of the normalized amplitude with respect
to the centre of the primary beam. For the two pointings adjacent to the
nucleus, we subtracted the nucleus in each ``snapshot'' of short integration
time to remove instrumental effects as the nucleus moves through the primary
beam. This procedure gave us an r.m.s.\ noise level of $16\,\mu\rm Jy
beam^{-1}$ in Stokes $Q$ and $U$, in agreement with the theoretically expected
value, where the maps are 8largely free of spurious emission from the nucleus.
The $(u,v)$-plane is not filled below a baseline of 30\,m corresponding to a
$(u,v)$-distance of $1\,\rm k\lambda$ at $\lambda$3\,cm. Therefore, any
structures with an angular diameter larger than $3\arcmin$ cannot be
imaged. To reconstruct the large angular scales, we merged the VLA and
Effelsberg data using {\tt IMERG}, which is part of AIPS. The Effelsberg maps
were presented in the two preceding papers of this series.  The merging was
done by a Fourier transform of the two maps to be combined, where an overlap
in the $(u,v)$-plane had to be specified as {\tt uvrange}, so that the
observations could be matched. We compared the integrated flux densities in
several separated regions in the Stokes $I$ map both, in the disc and
halo. The flux densities agreed within $10\%$ in each of the regions for the
chosen overlap of $0-1.2\rm k\lambda$. The $Q$ and $U$ maps were merged using
the same optimum uvrange as found for $I$.
\subsection{VLA archive data}
Archive observations at $\lambda$6\,cm in the C-configuration are available that have
not been published so far. They consist of five pointings along the major axis with
30\,min total integration time each, but spread over 6\,h to have a good coverage in
the $(u,v)$-plane. The data were calibrated as described for the $\lambda$3\,cm
observations in Stokes $I$ and polarization. As we only have one pointing
centred on the nucleus there are no sidelobes of spurious polarization. We
inverted the $(u,v)$-data using robust weighting ($\rm robust=0$) and cleaned
down to $2\times$ the r.m.s.\ noise level. The map was merged with Effelsberg
single-dish data using {\tt IMERG}, with an overlap of $0.2-0.8\rm k
\lambda$. The r.m.s.\ noise level in polarization is $40\,\rm \mu
  Jy\,beam^{-1}$, in agreement with what we would expect. In total power the
  r.m.s.\ noise level is $100\,\rm \mu Jy\,beam^{-1}$, limited by the high dynamic
  range of 8300 due to the bright nuclear source.
  Observations in the B-configuration at $\lambda$20\,cm were also taken from
  archive data. These were already presented in \citet{carilli_92a} albeit
  with lower resolution, because they were combined with a more compact
  configuration. We attempted a polarization calibration, but this was not
  possible, indicated by a large variation in the polarization angle of the
  primary calibrator with baseline and time. For a successful polarization
  calibration, the variation should only be a few degrees at most. The
  observations were centred on the nucleus, where one pointing is enough to
  cover the central part of the galaxy. We combined the $\lambda$20\,cm
  observations in B-configuration in total intensity with those in
  C-configuration, so that we are sensitive to angular scales of up to
  $15\arcmin$, i.e.\ to all but the largest angular scales. The r.m.s.\ noise
  level is, with $400\,\rm\mu Jy\,beam^{-1}$, slightly higher than in the map
  presented by \citet{carilli_92a}, which has a noise level of $350\,\rm\mu
  Jy\,beam^{-1}$ at $30\arcsec$ resolution (Paper~I).
  Furthermore, there is some high-resolution data at $\lambda$20\,cm in VLA
  A-configuration that was presented in \citet{ulvestad_97a}, although they
  did not study the extended emission. The data were calibrated in the
  standard fashion in AIPS where we used a model for the primary calibrator
  3C48, because it is resolved on the longest baselines. We cleaned the
  $(u,v)$-data using robust weighting (robust=0) to 2$\times$ the r.m.s.\
  noise level. The final noise level is $50\,\mu \rm Jy/beam$ at a resolution
  of $2\farcs 2\times 1\farcs 3$ ($p.a.$=$-2\degr$), which agrees with
  \citet{ulvestad_97a}. We attempted a polarization calibration of the data,
  but it failed again, as indicated by the large variation in the primary
  calibrator's polarization angle.  The details of the maps presented in this
  paper are summarized in Table\,\ref{tab:VLA}.
  For the analysis of the rotation measure we also created natural weighted
  maps at $\lambda\lambda$ 6\,cm and 3\,cm with $10\farcs 5$ resolution. At
  $\lambda$6\,cm we combined the data in polarization from our previously
  published observations in D-configuration with those in C-configuration. At
  both wavelengths the maps were merged with Effelsberg single-dish data.
\section{Morphology of the radio continuum emission}
\label{sec:structure_of_the_radio_continuum_emission}
\subsection{Nuclear region and bar}
\label{subsec:bar}
\begin{figure}[tbph]
\resizebox{\hsize}{!}{ \includegraphics[angle=270]{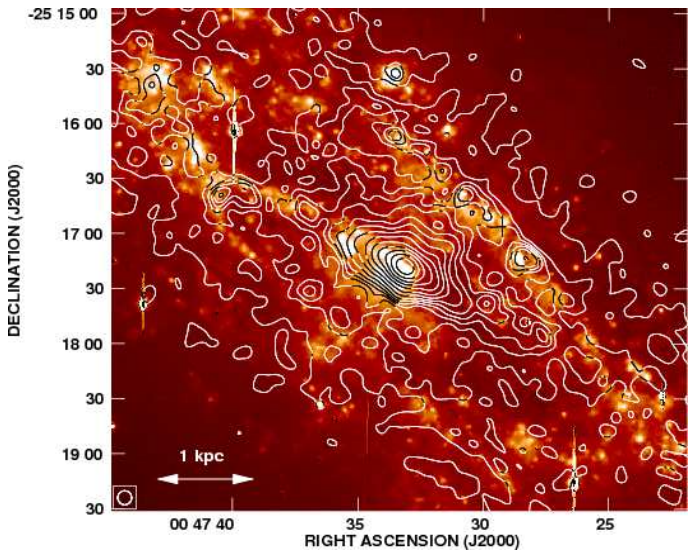}}
\caption{Total power radio continuum at $\lambda 20\,{\rm cm}$ with
  $7\farcs 5$ resolution. Contours are at 3, 6, 8, 10, 12, 15, 20, 30, 50, 100,
  200, 400, 800, and 1600 $\times$ 400\,$\mu$Jy/beam. The background is
  the H$\alpha$ image from \citet{hoopes_96a}.}
\label{fig:20cm_tp}
\end{figure}

\begin{figure}[tbph]
\resizebox{\hsize}{!}{ \includegraphics[angle=270]{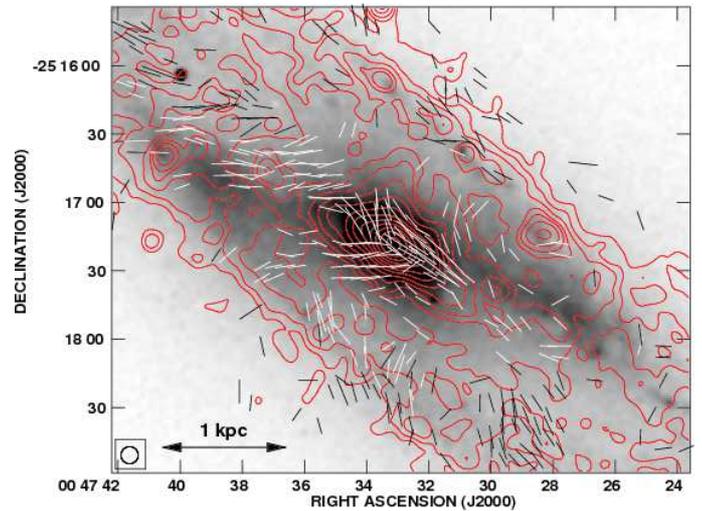}}
\caption{Total power radio continuum at $\lambda 6\,{\rm cm}$ with $7\farcs 5$
  resolution. Contours are at 5, 7, 10, 15, 20, 30, 50, 100, 200, 400, 800, 1600,
  3200, and 6400 $\times$ 100\,$\mu$Jy/beam. The vectors show the orientation
  of the magnetic field (not corrected for Faraday rotation) where a vector
  length of $1\arcsec$ is equivalent to 12.5\,$\mu$Jy/beam of polarized
  intensity. The background is a $K$ band image from the 2MASS survey
  \citep{jarrett_03a}.}
\label{fig:6cm_tpa}
\end{figure}

In Fig.\,\ref{fig:20cm_tp} we show the $\lambda$20\,cm total power radio
continuum distribution overlaid on an H$\alpha$ image by
\citet{hoopes_96a}. The radio continuum (though mostly non-thermal) traces the
\ion{H}{II} regions well, which are visible as clumps in the H$\alpha$
image. They form a ring around the centre, where the spiral arm NW of the
nucleus is prominent. We discover a long structure in the inner region,
extending both NE and SW of the nucleus, resembling a radio continuum bar that
has so far not been reported for this galaxy.  This radio bar has a position
angle of $63\degr \pm 1\degr$. In Fig.\,\ref{fig:6cm_tpa} we overlayed the
$\lambda$6\,cm continuum emission onto a 2MASS $K$-band image that shows the
stellar bar \citep{jarrett_03a}. The radio and the stellar bar have about the
same length, and they align well, but the former is shifted clockwise with respect
to the latter (the stellar bar has a position angle of $70\degr\pm 1\degr$).
The galaxy rotates clockwise with trailing spiral arms.  According to HI
observations in the barred galaxies NGC\,1097 and NGC\,1365 bar patterns
rotate more slowly than the gas \citep{ondrechen_89b,ondrechen_89a}, so that
the gas flow enters the bar from behind; i.e.\ the bar is overtaken by the
gas. Therefore the radio bar is \emph{downstream} of the stellar bar and on
the leading edge with respect to the rotation. This is the same behaviour as
in the prominent barred galaxy NGC\,1365 that has been extensively studied in
radio continuum \citep{beck_05b}. Gas and dust usually accumulates in the
downstream region of bars \citep{athanassoula_92a}.  \citet{sorai_00a} indeed
find that the molecular gas in NGC\,253 is located in the downstream region,
at $\rm p.a.=66\degr\pm 1\degr$, so that it is located between the stellar and
the radio bar. The inner part of the CO-bar of about $1\arcmin$ (1.1\,kpc)
length is also visible in the map by \citet{sakamoto_06a}
(Fig.\,\ref{fig:cm20tpCO12}).

The concentration of gas and dust may be due to a shock in the bar's potential
that also compresses the magnetic field. Compression likely plays a role at
least in the eastern bar because there the polarization degree is $22\pm7\%$
(at $\lambda$3\,cm), a high value for a location near the midplane.  The
sequence from downstream to upstream is therefore regular magnetic fields,
turbulent magnetic field, molecular gas, and stars. This is the reversed order
that was found in the arms of spiral galaxies, as in M51 by
\citet{patrikeev_06a}, so it warrants further investigation.
%
\begin{figure}
\resizebox{\hsize}{!}{ \includegraphics[angle=270,clip=]{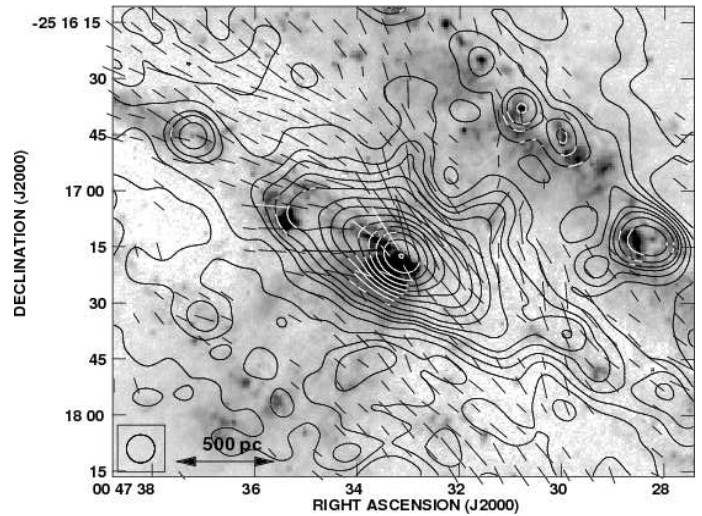}}
\caption{Total power radio continuum at $\lambda 3\,{\rm cm}$ at $7\farcs 5$
  resolution. Contours are at 3, 5, 7, 10, 15, 20, 30, 50, 100, 200, 400, 800,
  1600, 3200, and 6400 $\times$ 50\,$\mu$Jy/beam. Vectors show the orientation
  of the magnetic field (assuming negligible Faraday rotation) where $1\arcsec$
  vector length is equivalent to $12.5\mu\rm Jy\,beam^{-1}$ polarized
  intensity. The background is the H$\alpha$ image from
  \citet{westmoquette_11a}, which was scaled with a square-root function to
  enhance weak emission.}
\label{fig:3cm_tpa}
\end{figure}
\subsection{Filamentary emission}
\label{subsec:filaments}
Other distinctive morphological features are the two horn-like structures
that are sticking out of the central region roughly perpendicular to the
direction of the bar, each consisting of two filaments\footnote{We note that
  the observed filaments are apparent structures observed in
  projection. They will be explained later as the walls of the nuclear outflow
  cones.}. The northern one can be clearly seen in the $\lambda$3\,cm map that
we present in Fig.\,\ref{fig:3cm_tpa} with a length of $30\arcsec$ = $600\,\rm
pc$. Both structures are marked in Fig.\,\ref{fig:cm20tpCO12} as F1 and F2
north and as F3 and F4 south of the galactic plane. They are emerging from the
central region that contains the nucleus, which is a strong total power
source. The nuclear region is extended in the direction of the bar and has a
length of $1200\,\rm pc$ at this resolution.  The nuclear source itself is not
resolved at $7\farcs 5$ (150\,pc) angular resolution. In the optical regime
the morphology is dominated by absorption owing to dust filaments as shown in
the high-resolution HST image from the ANGST survey \citep{dalcaton_07a}
(Fig.\,\ref{fig:3cm_hst}).
\begin{figure}[tbhp]
\resizebox{\hsize}{!}{\includegraphics[angle=270]{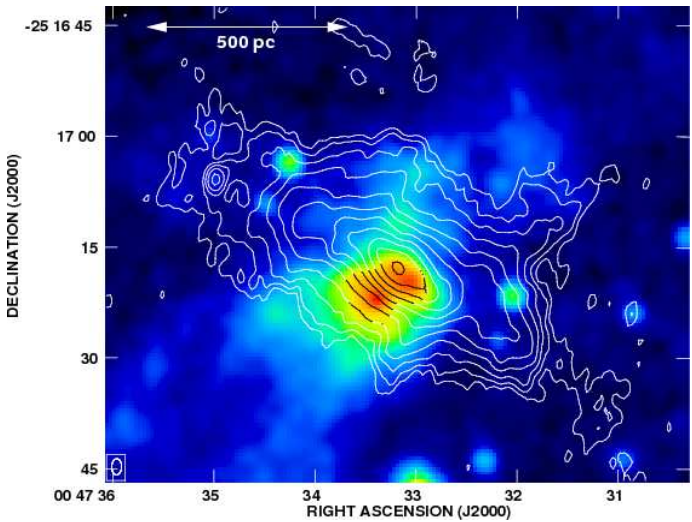}}
\caption{Total power radio continuum at $\lambda 20\,{\rm cm}$ with $1\farcs 3\times 2\farcs 2$ resolution. Contours
  are at $-6$, $-3$, 3, 6, 12, 25, 50, 100, 200, 400, 800, 1600, and 3200
  $\times$ $50\,\mu\rm Jy\,beam^{-1}$. The background is a map of the X-ray
  emission between 0.5 and 5\,{\rm keV} measured with the \emph{Chandra} satellite
  convolved with a Gaussian to $2\arcsec$ resolution (M.\ Hardcastle, priv.\ com.).}
\label{fig:20cm_xray}
\end{figure}
The horn-like structures are also visible in the $\lambda$20\,cm
A-configuration map at a high resolution of $2\farcs 2\times1 \farcs 3$, shown
in Fig.\,\ref{fig:20cm_xray} as an overlay on X-ray emission from
\emph{Chandra} data (M.\ Hardcastle, priv.\ com.). At this high resolution the
nucleus (R.A.\ $00^{\rm h} 47^{\rm m} 33\fs 12$, Dec.\ $-25\degr 17\arcmin
17\farcs 3$) is resolved. The nuclear region now shows up as a structure of
$900 \times 600\,\rm pc$ in extent, with the longer extension aligned with the
molecular bar, rather than with the major axis. Noteworthy is the apparent
anti-correlation between the X-ray and the radio continuum emission in
locations where the X-ray emission traces an outflow of hot gas. This is
apparent in the thin X-ray filament (in green, R.A.\ $00^{\rm h} 47^{\rm m}
33\fs 2$, Dec.\ $-25\degr 17\arcmin 10\arcsec$) pointing to the N and in the
SE outflow cone, where the radio contours bend slightly inwards. The decrease
in scaleheight indicates enhanced synchrotron loss due to the strong magnetic
field in the outflow cone (see Paper~II for details).
The filamentary structures are better visible in the grey-scale map presented
in Fig.\,\ref{fig:cm20tpCO12}, where the contour lines show $^{12}$CO J=2-1
emission observed with the Submillimeter Array \citep{sakamoto_06a}. There are
four filaments that all have widths of about one beam diameter ($40\,\rm pc$),
so that their true width is smaller and have a length of up to $400\,\rm
pc$. The filaments are not exactly perpendicular to the extension but are
slightly inclined, forming rather the boundary of a cone-like structure. The
position angle of the filaments are $-53\degr\pm 1\degr$ (F1), $-26\degr \pm
1\degr$ (F2), $141\degr \pm 2\degr$ (F3), and $116\degr \pm 4\degr$ (F4). The
opening angles of the cones (between F1 and F2) are therefore $27\degr\pm
1\degr$ in the NW and between F3 and F4 $25\degr\pm 4\degr$ in the SE,
respectively.
\begin{figure}[tbhp]
\resizebox{\hsize}{!}{ \includegraphics{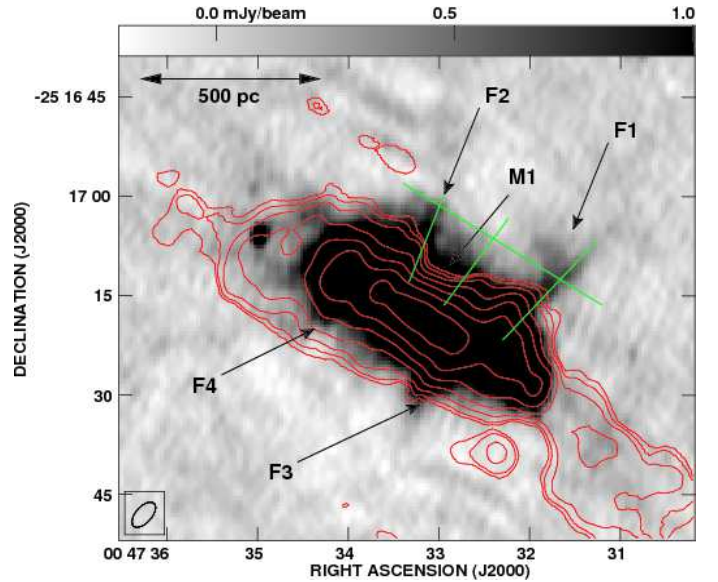}}
\caption{$^{12}$CO J=2-1 emission from observations with the Submillimeter
  Array at $4\farcs 6\times 2\farcs 4$ resolution ($\rm p.a. = -41\fdg 2$)
  from \citet{sakamoto_06a}. Contours are at 3, 6, 12, 25, 50, 100, and 200
  $\times$ $7\,\rm Jy\,beam^{-1}\,km\,s^{-1}$. The background image is the
  $\lambda$20\,cm total power emission with $1\farcs 3\times 2\farcs 2$ resolution
  as shown in Fig.\,\ref{fig:20cm_xray},
  which shows the weak emission. The green lines show the location of the
  profiles discussed in Sect.\,\ref{subsec:emission_profiles}. The four
  filaments F1--4 and the minimum M1 are denoted.}
\label{fig:cm20tpCO12}
\end{figure}
\setcounter{figure}{05}
\begin{figure}[tbhp]
\resizebox{\hsize}{!}{\includegraphics[angle=270]{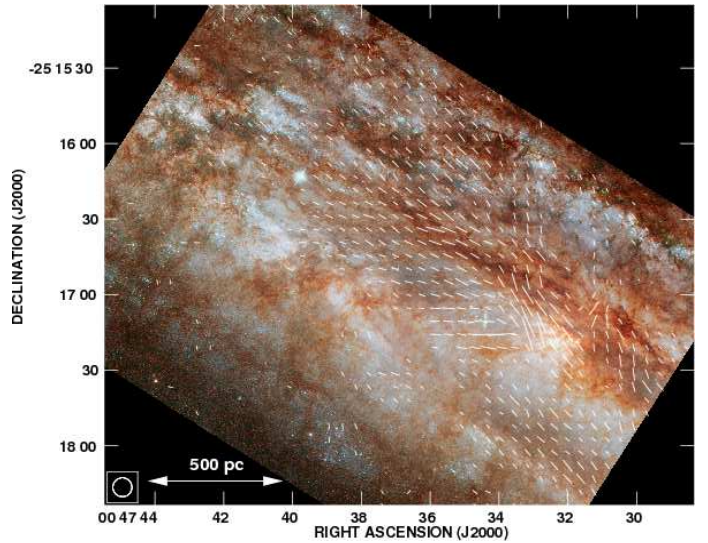}}
\caption{Magnetic field orientation at $\lambda$3\,cm with 7\farcs 5
  resolution overlaid on an HST image from the ANGST survey
  \citep{dalcaton_07a}. A vector length of $1\arcsec$ is equivalent to a
  polarized intensity of $21.3\,\mu\rm Jy\,beam^{-1}$. The image shows
  only the centre and the NE part.}
\label{fig:3cm_hst}
\end{figure}
\subsection{Outflow cones}
\label{subsec:cones}
The nuclear outflow of the starburst galaxy is visible as plumes of H$\alpha$
and soft X-ray emission. In the NW on the far side of the halo, the emission
is heavily absorbed, so that the outflow is much less visible at these
wavelengths. The H$\alpha$ filaments bordering the SE outflow cone coincide in
shape and position with the radio filaments F3 and F4. The NW outflow cone is
seen in H$\alpha$ only as a weak triangular shape, where the border again
coincides with the radio filaments, here F1 and F2. In Fig.\,\ref{fig:3cm_tpa}
the H$\alpha$ mission can be understood as limb brightening of the cone
\citep[see also Fig.\,2 in][]{westmoquette_11a}. The H$\alpha$ emission is
presumably material around the rim of the cone, which cooled from the hot gas
in the inner part of the cone with a temperature of $10^7\,\rm K$.  The
filaments F1 and F2 in the NW are not visible in H$\alpha$, probably due to
absorption. The soft X-ray emission also shows some filamentary emission at
the boundary of the bubbles interpreted as cooled material
\citep{strickland_00a}. Using more sensitive XMM observations albeit with
coarse resolution, \citet{bauer_07a} report detecting the actually hot plasma
of the outflow.
By comparing the radio contour lines in Fig.\,\ref{fig:20cm_xray} with
the CO contour lines in Fig.\,\ref{fig:cm20tpCO12}, we can see
several similarities. The radio filaments agree in position with
extensions in CO emission except at the weak filament F4 that is not
prominent in CO. The minimum M1 on the northern side of the radio is
also pronounced in CO. The similarities in the contour lines hold
also for the shape for part of the nuclear region. The obvious
exceptions are extensions in CO east and southwest of the nuclear
region. We will attempt a more quantitative comparison between the
radio and CO emission in Sect.\,\ref{subsec:star_formation}.
\setcounter{figure}{6}
\begin{figure*}[tbhp]
\resizebox{\hsize}{!}{ \includegraphics{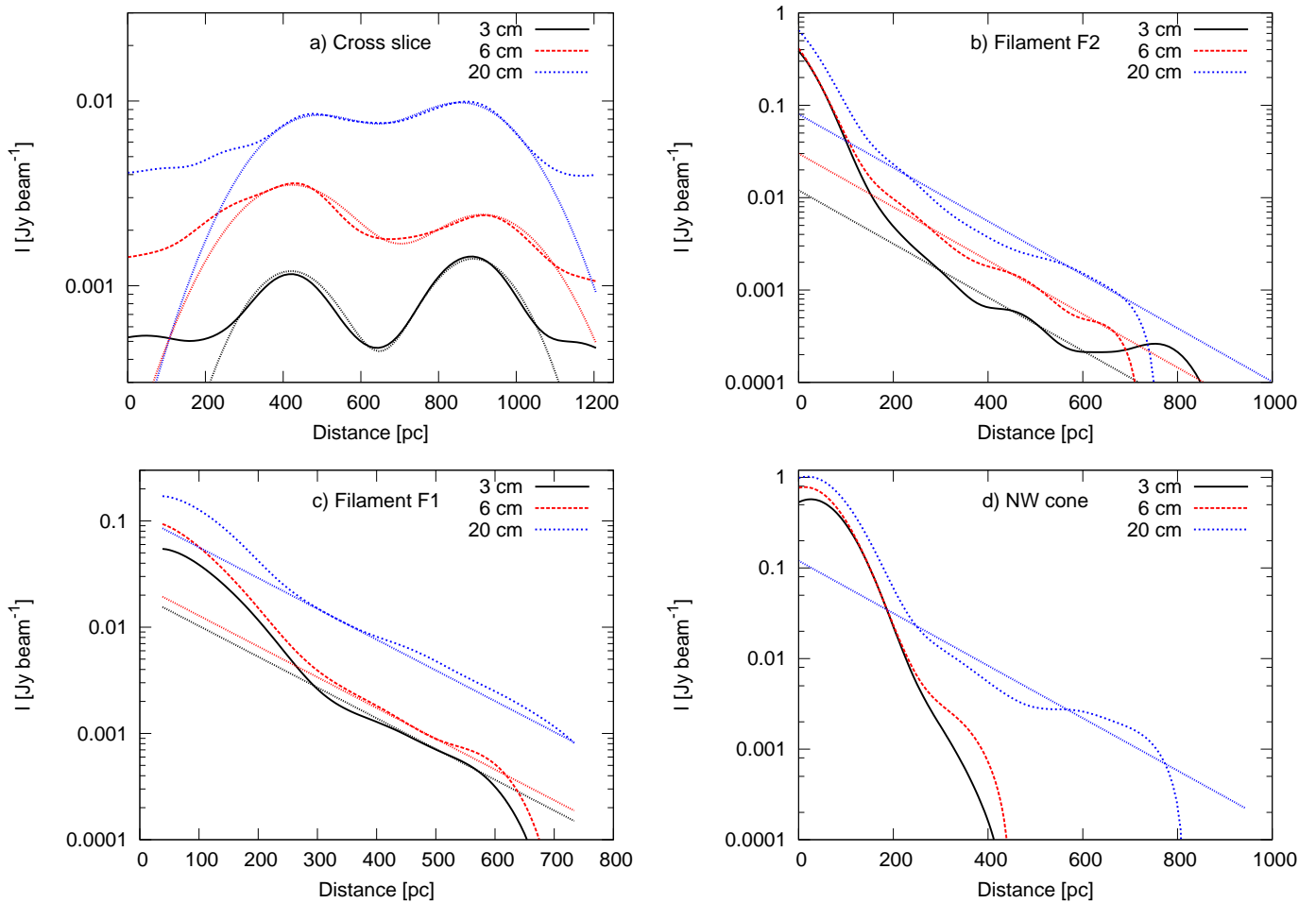}}
\caption{Slices of continuum emission where the dotted lines show the measured
  profiles and the solid lines the fit to the data. a) Slice across the filaments
  parallel to the central region at a distance of 300\,pc. b) and c) Slices
  along the filaments F2 and F1, respectively. d) Slice along the NW outflow
  cone. For slices (b)-(d) a constant level of background emission estimated
  from (a) was subtracted.}
\label{fig:slices}
\end{figure*}
\subsection{Emission profiles}
\label{subsec:emission_profiles}
A useful way to characterize the distribution of the continuum emission is to
measure scaleheights. We created profiles of the radio continuum emission with
a fixed resolution of $7\farcs 5 = 150\,\rm pc$ for the filaments F1 and
F2. The profiles, along with exponential fits, are presented in
Fig.\,\ref{fig:slices}, where we used a slice width of $2\arcsec$, which is
smaller than the resolution. Also, we used a perpendicular profile across the
filaments, which shows the filaments and the depression between them. For the
slices along the filaments and the cone (Fig.\,\ref{fig:slices}b-d), we
subtracted a constant level of background emission estimated from the cross
slice (Fig.\,\ref{fig:slices}a). The contrast between the filaments and the
depression depends on wavelength. The depression is most prominent at
$\lambda$3\,cm, with the level of emission in the outflow cone decreasing to
the surrounding level. Therefore, at $\lambda$3\,cm we do not detect any
continuum emission coming from the interior of the outflow cone. At
$\lambda$6\,cm the intensity inside the outflow cone is not entirely
decreasing to the surrounding level and at $\lambda$20\,cm the depression is
only shallow. See also Figs.\,\ref{fig:20cm_tp}-\ref{fig:3cm_tpa}, where the
appearance of the filaments clearly depends on wavelength.
We fitted exponential functions to the emission profiles to measure the
scaleheights. We found a scaleheight of $150\pm20\,\rm pc$ in both filaments
at all three wavelengths as shown in Figs.\,\ref{fig:slices}b and c. This is
remarkable because the scaleheights of the continuum emission in the halo
increase with increasing wavelength (Paper~II). This is understood by the
increase in the lifetime of cosmic-ray electrons with increasing
wavelength. The wavelength-independent continuum scaleheight of the filaments
indicates that we see a decrease in the magnetic field strength, rather than
that of the cosmic-ray electrons density. 
We fitted Gaussian profiles to the cross slices in Fig.\,\ref{fig:slices}a to
measure the width of the filaments as a function of wavelength. Under the
assumption that the filaments have a Gaussian profile the observed width can
be expressed as $\rm FWHM^2_{\rm obs} = FWHM^2_{\rm fil} + (150\,pc)^2$ to
account for the limited resolution, which is comparable to the width of the
filaments.  We found FWHMs of $330\pm 15\,\rm pc$ ($\lambda 20\,\rm cm$),
$350\pm 15\,\rm pc$ ($\lambda 6\,\rm cm$), and $250\pm 15\,\rm pc$ ($\lambda
3\,\rm cm$). We note that these values are considerably higher than the
filament width of $< 40\,\rm pc$ that we inferred from the $\lambda$20\,cm
high-resolution map (Sect.\,\ref{subsec:filaments}). This result suggests a
weak, extended radio emission component associated with the filaments that is
not detected in the high-resolution map with a reduced sensitivity to extended
emission. Both the magnetic field and cosmic-ray electrons can contribute to
this weak component.
\begin{figure}[tbhp]
\resizebox{\hsize}{!}{\includegraphics{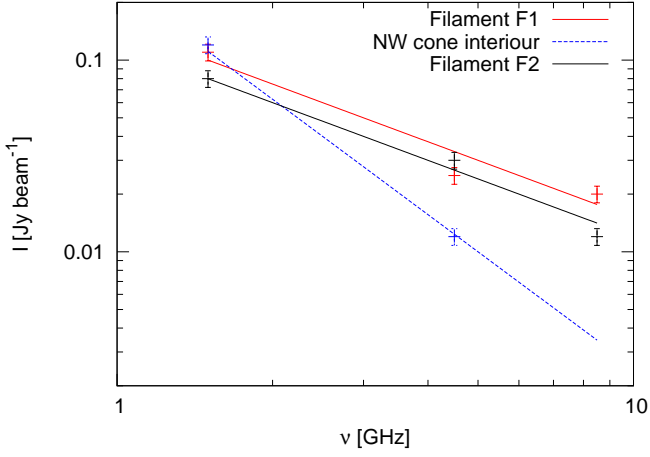}}
\caption{Spectra of the radio continuum flux density as measured from the fits
  to the data shown in Fig.\,\ref{fig:slices}.}
\label{fig:flux}
\end{figure}
The frequency dependence of the filament width suggests a transport effect of
the cosmic rays (see Sect.\,\ref{subsec:cr_transport}). A wavelength-dependent
filament width is also found in supernova remnants, where the synchrotron
emission in the X-ray regime exhibits extremely thin filaments ($<0.1\,\rm
pc$) with strong magnetic fields, whereas the radio filaments are much broader
as the result of the longer lifetime and diffusion of the lower energy
electrons \citep{reynolds_11a}.
\subsection{Radio spectral index}
\label{subsec:spix}
The emission profiles with the same scaleheight in the filaments imply that
the radio spectral index $\alpha$ (defined as $S\propto \nu^\alpha$) is
constant with respect to the distance from the disc.  To calculate the radio
spectral index, we took the flux density from the fits shown in
Fig.\,\ref{fig:slices} extrapolated to a distance of 0\,pc. For both filaments
we find $\alpha = -1.0\pm0.1$, with no significant curvature in the spectra
shown in Fig.\,\ref{fig:flux}. In filament F2 the $\lambda$3\,cm flux density
is low hinting at strong electron losses. In contrast, the $\lambda$3\,cm flux
density is high in filament F1 where the thermal fraction may be greater. Of
course, with only three different wavelengths, we cannot measure any curvature
precisely but can assume a linear spectrum. The radio spectra in the filaments
are consistent with a non-thermal spectrum within the error interval, with a
non-thermal radio spectral index of $\alpha_{\rm nt}=-1$. In the interior of
the cone (see Fig.\,\ref{fig:slices}a), the radio spectral index is $-2.0\pm
0.2$ between $\lambda\lambda$ 20 and 6\,cm, indicating strong energy losses of
the electrons.
\begin{figure}[tbph]
\resizebox{\hsize}{!}{ \includegraphics{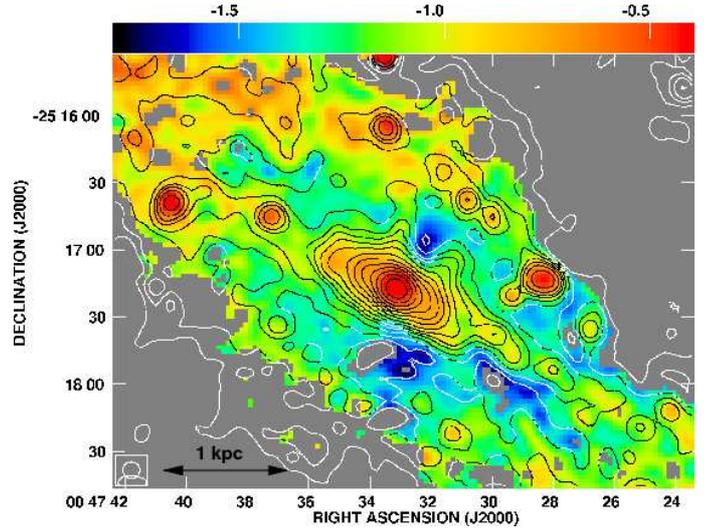}}
\caption{Radio spectral index between $\lambda\lambda$ 20 and 3\,cm at
  $7\farcs 5$ resolution as grey-scale. Contours show the continuum emission at
$\lambda$3\,cm as in Fig.\,\ref{fig:3cm_tpa}.}
\label{fig:20_3_sp}
\end{figure}
\begin{figure}[tbhp]
\resizebox{\hsize}{!}{ \includegraphics{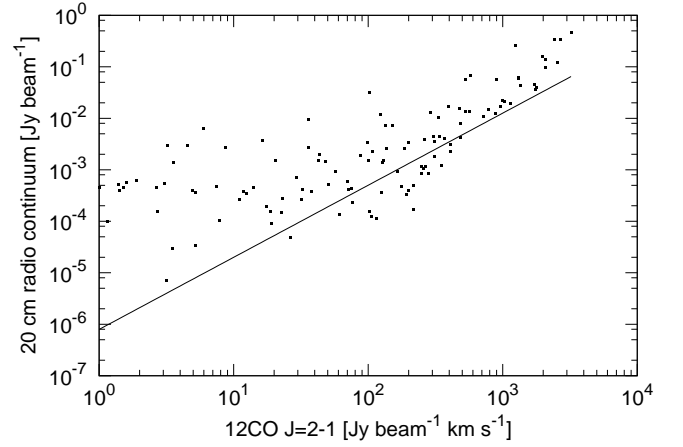}}
\caption{Pixel-by-pixel comparison of the $\lambda 20\,\rm cm$ radio continuum
  intensity with the $^{12}$CO J=2-1 intensity at 88\,pc resolution. The line
  shows the theoretical expectation using the conversion into star formation
  rates as explained in the text.}
\label{fig:n253_cr}
\end{figure}
In Fig.\,\ref{fig:20_3_sp} we present the distribution of the radio spectral
index between $\lambda\lambda$ 20 and 3\,cm. We clipped each map below
4$\times$ the r.m.s.\ noise level prior to the combination, so that the
maximum error is $\pm0.3$ in the radio spectral index. The spectrum is flat in
the centre around the nucleus but steepens significantly away from it. The two
filaments in the NW and the radio bar show up as extensions of the flat
central region. The flat spectral index of the central region also extends
into the radio bar on both sides of the nuclear region.  At the location of
the NW outflow cone (R.A.\ $00^{\rm h} 47^{\rm m} 32^{\rm s}$, Dec.\ $-25\degr
16\arcmin 50\arcsec$), the radio spectral index steepens significantly, as
expected for strong radiation losses of the electrons
(Sect.\,\ref{subsec:cr_transport}). We checked carefully that this feature is
not caused by the merging of the interferometric data with the single-dish
data. It is already apparent in the radio spectral index map constructed from
VLA data alone. The steep region south of the nucleus (R.A.\ $00^{\rm h}
47^{\rm m} 33^{\rm s}$, Dec.\ $-25\degr 17\arcmin 50\arcsec$), however, is
less certain. At $\lambda$3\,cm we observe a minimum in the total power
emission that may be caused by the high dynamic range close to the
nucleus. Also the position does not not agree with the SE outflow cone (see
Fig.\,\ref{fig:3cm_tpa}), so that this feature could be an artefact.
\setcounter{figure}{10}
\begin{figure}[tbhp]
\resizebox{\hsize}{!}{
\includegraphics[angle=270]{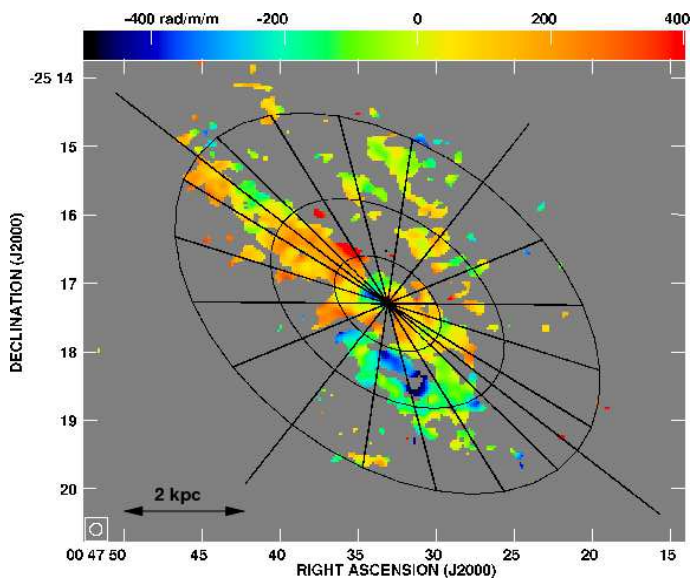}} \caption{Rotation
measure distribution between $\lambda\lambda$ 6 and 3\,cm with a
resolution of $10\farcs 5$. The major
and minor axes and the position of the sectors are shown.}
\label{fig:rm6_3}
\end{figure}
\subsection{Radio continuum as star formation tracer}
\label{subsec:star_formation}
The radio continuum emission is a star formation tracer because it
is sensitive to the thermal and the non-thermal emission. UV-photons
emitted by massive stars ionize the hydrogen that can be seen as
thermal H$\alpha$ and free-free radio emission. Supernova
remnants that mark the end of life of massive stars accelerate
cosmic-ray electrons that radiate synchrotron emission, the
non-thermal component of the radio emission. \citet{condon_92a}
calculated the star formation rate or, for resolved emission, the
star formation rate density as
\begin{equation}
\left (\frac{\rm SFRD}{\rm M_\odot\, yr^{-1}\, kpc^{-2}}\right )_{>0.1\,M_\odot} =
1.4\times10^{-7} \left ( \frac{I_{20\,\rm cm}}{\rm
    Jy\,ster^{-1}}\right ),
\label{eq:sfrd_r}
\end{equation}
where $I_{20\,\rm cm}$ is the radio continuum intensity at $\lambda$20\,cm.
Another star formation tracer is the molecular hydrogen traced by
the $^{12}$CO J=2-1 emission. According to \citet{kennicutt_98a} the
star formation rate density relates to the mass surface density of combined atomic and molecular hydrogen $\Sigma_{\rm H2}$ as
\begin{equation}
\rm \left (\frac{SFRD}{\rm M_\odot\, yr^{-1}\, kpc^{-2}}\right )  = 2.5
\times 10^{-10} \left (\frac{\Sigma_{\rm gas}}{\rm M_\odot\, yr^{-1}\,
    kpc^{-2}} \right )^{1.4}.
\label{eq:sfrd_c}
\end{equation}
We assume that the entire gas in the nuclear region is of molecular nature and
estimate the gas mass thus from the molecular hydrogen alone, which we derived
from the CO map.  We adopted a ratio between $^{12}$CO(2-1) and $^{12}$CO(1-0)
of 0.9 found by \citet{weiss_05a} for the starburst galaxy M\,82. Also we used
the conversion of $5\times 10^{19}$ molecules $\rm cm^{-2} / (K\, km\,s^{-1})$
to convert from CO to $\rm H_2$ column density, which was derived by
\citet{downes_98a} for ultraluminous far-infrared (FIR) galaxies and is widely
used for starburst galaxies in the literature.
\begin{figure}[tbhp]
\resizebox{\hsize}{!}{
\includegraphics[angle=270]{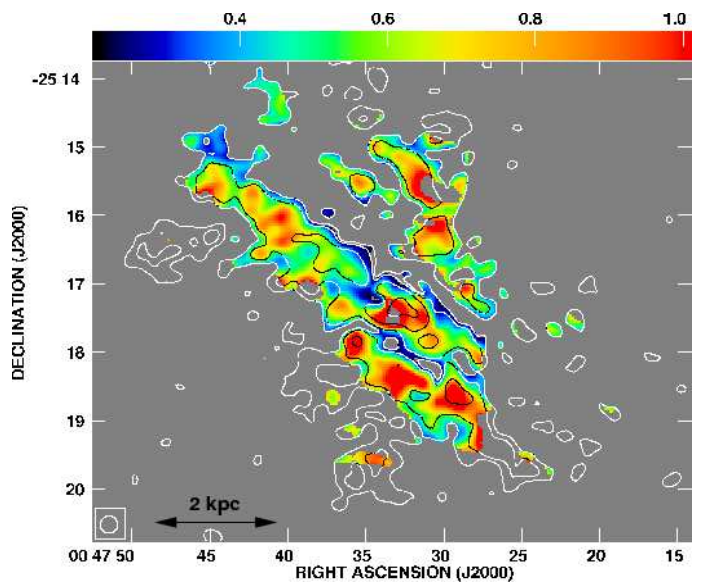}} \caption{Distribution of the
depolarization between $\lambda\lambda$ 6 and 3\,cm at a resolution of
$15\arcsec$. Contours show the distribution of the polarized emission at
$\lambda\rm 6\,cm$ with contours at 3, 5, 10, and 20 $\times$ $70\,\mu\rm Jy\,
beam^{-1}$, the r.m.s.\ noise level.} 
\label{fig:dep}
\end{figure}
In Fig.\,\ref{fig:n253_cr} we compare the radio continuum intensity at
$\lambda 20\,\rm cm$ and the $^{12}$CO J=2-1 intensity on a scale of $4\farcs
6 = 88\,\rm pc$ in a pixel-by-pixel plot. The line shows the relation, if we
use the conversion into star-formation rate densities. The scatter in the data
points is large, with the radio flux densities on average a factor of 4
brighter than theoretically expected. \citet{ott_05a} find that the integrated
star formation rate in the nucleus of NGC\,253 as derived from
$\lambda$1.2\,cm radio continuum observations agrees with that of the FIR,
suggesting that the radio emission is a reliable tracer for star
formation. Recent studies established such a correlation in spatially resolved
observations. \citet{dumas_11a} confirm a tight correlation between radio
emission and various other star formation tracers down to a scale of
240\,pc. We do not find such a tight correlation in the nucleus of NGC\,253. A
possible explanation are transport effects of the cosmic rays in the powerful
nuclear outflow, which we discuss in
Sect.\,\ref{subsec:cr_transport}--\ref{subsec:radio-fir}. If cosmic rays
diffuse away from the star formation sites, the radio emission gets decoupled
from other star formation tracers, and the relation between star formation rate
and radio emission breaks down.
\section{Magnetic field structure}
\label{sec:magnetic_field_structure}
\subsection{Magnetic field orientation}
In Figs.\,\ref{fig:3cm_tpa} and \ref{fig:3cm_hst} we show the magnetic field
orientation at $\lambda$3\,cm, where the length of the vectors is proportional
to the polarized intensity. The maximum polarized intensity is near the
nucleus, in the adjacent NE region. The distribution is asymmetric with
respect to the minor axis, the NE half having a higher fraction of
polarized intensity. There is a local minimum in the NW outflow cone where we
do not detect polarized emission, coincident with the minimum in total
power. The magnetic field orientation is parallel to the major axis only
NE of the nucleus near the stellar bar. In all other regions there is a
significant vertical component particularly in the two radio filaments
bordering the NW outflow cone. There, the magnetic field is also remarkably well
aligned with a polarization degree of $15\pm5\%$.
For the study of the Faraday rotation we made a map in polarization at
$\lambda$6\,cm (see Fig.\,\ref{fig:6cm_tpa}). Because the integration time is
short (30\,min), the map is not as sensitive as the $\lambda$3\,cm map. The
polarized emission is again asymmetric with respect to the minor axis with
most of the emission in the NE and is aligned roughly along the major
axis. The magnetic field orientation is different in places from that at
$\lambda$3\,cm indicating strong Faraday rotation, which we discuss in the
next section.
\subsection{Rotation measure distribution}
The rotation measure (RM) is given by the line-of-sight component of
the large-scale magnetic field. If the RM is positive the magnetic
field is pointing towards us and if it is negative it points away
from us. In Fig.\,\ref{fig:rm6_3} we present the RM distribution
between $\lambda\lambda$ 6 and 3\,cm at $10\farcs 5$ resolution. The
maps of the polarization angle were clipped below a polarized
intensity of 3$\times$ the r.m.s.\ noise level, prior to the
combination. The RM ambiguity is $\approx \pm n\cdot 1200\,\rm
rad\,m^{-2}$.
The RM changes on scales of a few beam diameters, corresponding to about
100\,pc. This finding agrees with the observations in face-on galaxies, which
show similar RM distributions, e.g. in M\,51 \citep{fletcher_11a}. There the
halo RM is also expected to contribute to the RM signal as a foreground to the
bright disc. Such frequent RM reversals can hardly be observed in edge-on
galaxies due to the effect of averaging along the line-of-sight.  The RM
fluctuations may be caused by field reversals on small scales, possibly an
``anisotropic'' field, which is generated from an isotropic turbulent field by
compression or shear\footnote{The observed B-vectors of linearly polarized
  emission can trace either \emph{regular} magnetic fields (i.e.\ preserving
  their direction within the telescope beam) or \emph{anisotropic} fields
  (i.e.\ with multiple field reversals within the beam). Anisotropic fields
  can be generated from turbulent fields by shear or compression. To
  distinguish between these two components, additional Faraday rotation data
  is needed. The fields observed in polarization are called ``ordered''
  throughout this paper.}.
The RM is typically between $-200$ and $+200\, \rm rad\, m^{-2}$, far away
from any RM ambiguity between $\lambda\lambda$ 6 and 3\,cm, so the
orientation of the magnetic field vectors at $\lambda 3\,\rm cm$ are rotated
by at most $\pm 15\degr$.  The rotation angle is small, so that we do not apply any
correction to the polarization angles at this wavelength. We note that the
galactic foreground RM for NGC\,253 ($l=97\degr$, $b=-88\degr$) is only a
small contribution at $10\pm 5\,\rm rad\,m^{-2}$
\citep{noutsos_08a,taylor_09a}, which we neglect in the further analysis.
\subsection{Faraday depolarization}
The observed degree of polarization is usually lower than the theoretically
highest value of about $ 74 \% $. Relevant depolarization may be caused by
wavelength-independent beam depolarization and by wavelength-dependent
depolarization effects, the Faraday depolarization. The latter consists of
differential Faraday rotation (related to the regular magnetic field strength
and thermal electron density within the emitting source, the galaxy), of
Faraday dispersion related to the turbulent magnetic field strength within the
emitting source (internal Faraday dispersion), and of a turbulent magnetic
field between the galaxy and the observer (external Faraday dispersion)
\citep{burn_66a,sokoloff_98a}. Usually, the differential Faraday rotation is
expected to contribute most to the observed depolarization within a spiral
galaxy, because its values are related to the observed RMs within this galaxy.
Observationally, the Faraday depolarization is calculated as the ratio between
the observed degrees of polarization in maps at two different wavelengths
smoothed to the same angular resolution or directly by $\rm DP = (PI_1/PI_2)
\times (\nu_2 / \nu_1)^{\alpha_{nt}}$, where $\alpha_{\rm nt}$ is the
non-thermal spectral index. Using $\alpha_{\rm nt} = -1$ as measured Paper~I,
we calculated DP between $\lambda\lambda$ 6 and 3\,cm at $15\arcsec$
resolution presented in Fig.\,\ref{fig:dep}.
The value of DP is for most parts in the galaxy is rather low with values
between $0.6$ and $1.0$ where $\rm |RM|$ reaches values up to $ 250 \,\rm
rad\,m^{-2}$. As a result, the expected depolarization at $\lambda 6\,\rm cm$ due to
differential Faraday rotation alone leads to $ 0.65 \le \rm DP \le 1.0 $ and
indicates that other depolarization effects are negligible. The depolarization
increases with higher distances N of the major axis up to values of about
$\rm RM= 350 \,\rm rad\,m^{-2}$ where the differential depolarization leads to
$ \rm DP = 0.2 $ , followed by a stripe with no polarized emission at $\lambda
6\,\rm cm$ parallel to the major axis but offset in a NW direction. This
stripe has an almost linear, well-defined boundary, and is probably caused by
even stronger Faraday depolarization of the regular and turbulent magnetic
field in the disc that lies in front.
More to the NW, we again detect polarized emission at $\lambda 6\,\rm cm$
where DP again lies within 0.6 and 1.0. We conclude that if we restrict the
study of RM to the near side (SE of the major axis), our analysis should
hardly be affected by depolarization.
\subsection{Disc magnetic field}
\begin{figure*}[tbhp]
\resizebox{\hsize}{!}{ \includegraphics{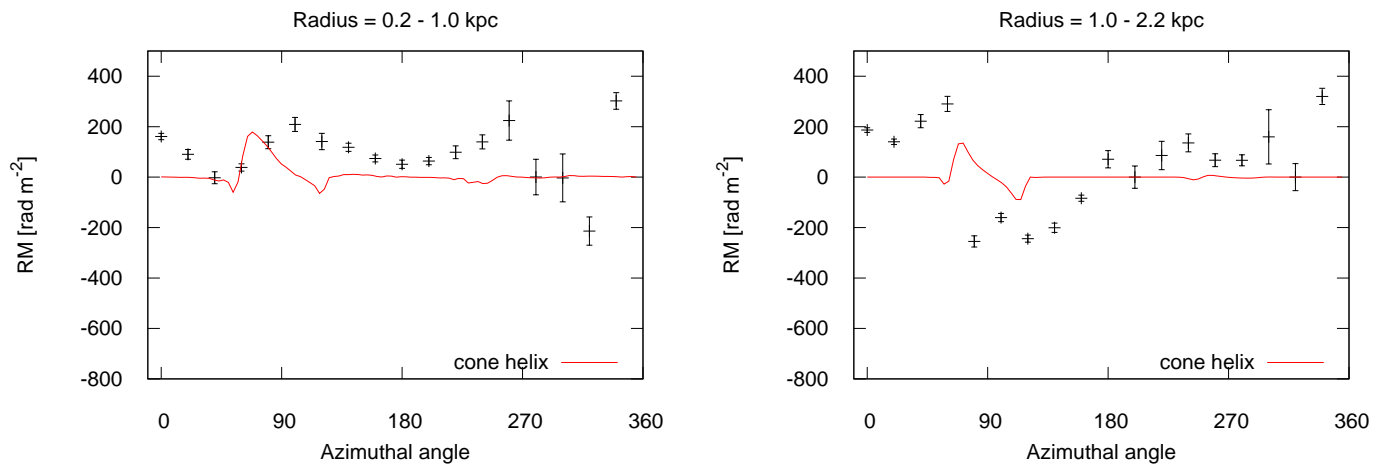}}
\caption{RMs averaged in azimuthal sectors in the inner disc. The
red line shows the model of the helical magnetic field in the
outflow cone as described in the text (no fit).}
\label{fig:radius}
\end{figure*}

According to the mean-field dynamo theory in thin-disc objects, the
axisymmetric spiral (ASS) mode ($m=0$) with even symmetry is the dominant
pattern of the regular field \citep{beck_96a}. Our azimuthal RM analysis in
Paper~II showed a single periodical variation, in accordance with an $m=0$
mode. Our new observations allowed us to test this in more detail and to
calculate refined ASS models. We compared the models with the observations
using averaging in sectors to plot the azimuthal RM variation. The radial
interval should avoid the starburst region and separate the molecular bar
region from the disc. The nuclear starburst region extends to a radius of
$\approx 300\,\rm pc$. The molecular bar extends to a galactocentric radius of
$3.5\,\rm kpc$ \citep{sorai_00a}, while the bright part extends to only about
1\,kpc (Fig.~\ref{fig:cm20tpCO12}). We chose 2.2\,kpc as a compromise. As the
outer radius we take 4.0\,kpc, up to where the signal-to-noise ratio allows us
the integration.  In summary, we used three radius intervals, $0.2-1.0\, \rm
kpc$, $1.0-2.2\, \rm kpc$ and $2.2-4.0\, \rm kpc$, where we chose an
inclination angle of $50\degr$ and an azimuthal spacing of $20\degr$ (the
position of the sector rings is shown in Fig.\,\ref{fig:rm6_3}). For the noise
estimate we used the r.m.s.\ noise values of Stokes $Q$ and $U$.
Figure~\ref{fig:radius} shows that, in the inner and middle ring,
$\rm |RM|$ has no maxima around the major axis (azimuthal angle $0\degr$
and $180\degr$), as would be expected for an axisymmetric
$m=0$ mode. Instead, $\rm |RM|$ is strongly enhanced and changes its
sign near the minor axis (azimuthal angles $90\degr$ and
$270\degr$) where the $m=0$ mode cannot contribute to RM. We conclude
that the disc magnetic field within $\approx 2.2\,\rm kpc$ does not
have a regular axisymmetric structure. To explain the lack of RM
near the major axis, we propose that the ordered field in the inner disc is
\emph{anisotropic}, i.e.\ it contributes to the polarized intensity
but has frequent reversals, so that it does not contribute to RM. In
Sect.\,\ref{subsec:rm_cone} we discuss a model of the conical
outflow that can explain the RM enhancements within $2.2\,\rm kpc$.
\begin{table}[t]
\caption[]{Parameters for the synthetic polarization maps (other parameters
  as in Paper~II).}
\vfill
\begin{center}
\begin{tabular}{lcr}
\hline\hline Parameter \T \B & Value & Notes\\\hline
Filament width \T & 40\,pc & from $\lambda$20\,cm\\
Base length  & 300\,pc & from $\lambda$20\,cm\\
Disc field scaleheight $h^{\rm disc}_{\rm B}$ & 1.6\,kpc & thin radio disc\\
Disc field strength $B^{\rm disc}_0$ & $4.4\,\mu\rm G$ & from equipartition\\
Filament field strength $B^{\rm fil}_0$ & $20\,\mu\rm G$ & from equipartition\\
Electron density $n_{\rm e}$ & $2\,\rm cm^{-3}$& from H$\alpha$ and soft
X-ray\\
Electron scaleheight $h_{\rm e}$ & 1400\,pc & from H$\alpha$\\
Cone opening angle $2\beta$ & $45\degr$ & \\
CRE scaleheight disc & 800\,pc & thin radio disc\\
CRE scaleheight filament & 200\,pc & filaments\\
Filament field scaleheight $h^{\rm fil}_{\rm B}$ \B & 600\,pc & from
equipartition \\
\hline \end{tabular}
\end{center}
\small{Note: CRE = cosmic-ray electrons} \label{tab:cone_parameters}
\end{table}
\begin{figure}[tbhp]
\resizebox{\hsize}{!}{ \includegraphics{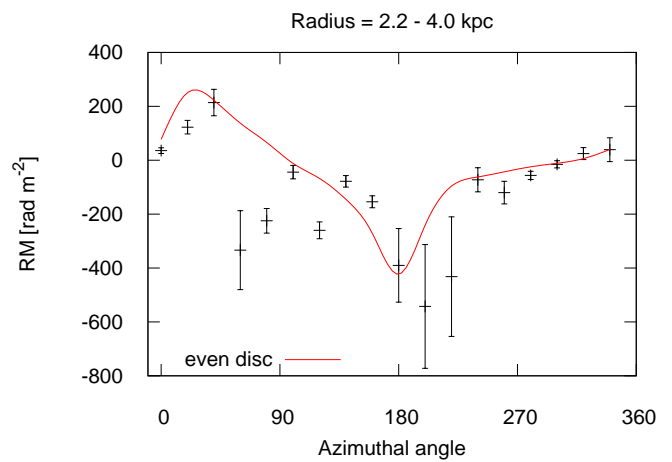}}
\caption{Azimuthally averaged RMs in the outer disc where the red
line shows an even-parity axisymmetric spiral model of the disc field.}
\label{fig:radius_22_40}
\end{figure}
In the outer ring ($2.2-4.0\,\rm kpc$), shown in
Fig.\,\ref{fig:radius_22_40}, the even-parity axisymmetric disc magnetic
field gives a reasonable fit to the data, except around $60\degr -
80\degr$. This deviation can be attributed to the ``radio spur''
that has already been identified by \citet{carilli_92a} in total power
radio continuum and which has a strong vertical field component.
This finding agrees with our previous observations (Paper~II),
where we detected an even-parity axisymmetric disc field at
lower resolution. Here we used the same model parameters as in Paper~II,
  except that the anisotropic component in the even disc model is zero. The
  model is therefore no fit to the data.
\subsection{Magnetic field in the outflow cone}
\label{subsec:rm_cone}
\begin{figure}[tbhp]
  \resizebox{\hsize}{!}{ \includegraphics{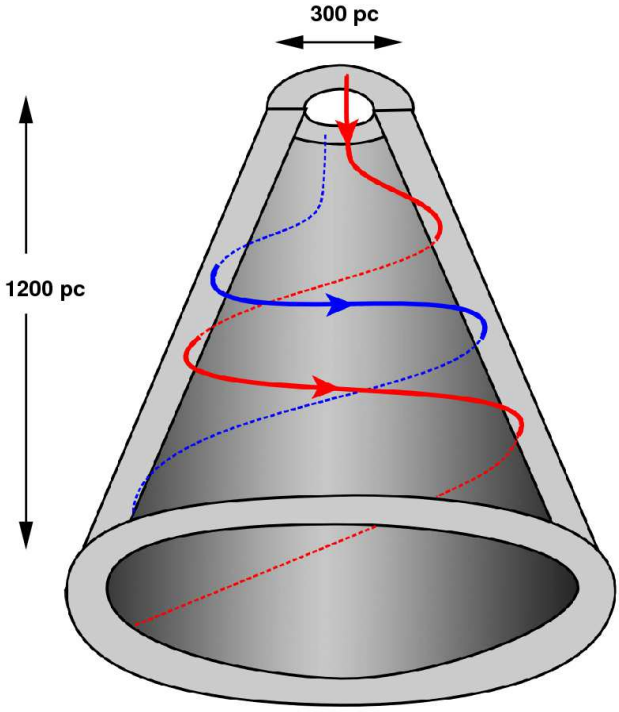}}
  \caption{Sketch of the magnetic field in the walls bordering the SE outflow
    cone, which appears as filaments when seen in projection. Solid lines show
    field lines on the front side of the cone, whereas dotted lines show them
    on the rear side.}
  \label{fig:fil}
\end{figure}

\begin{figure}[tbhp]
\resizebox{\hsize}{!}{
\includegraphics[angle=270]{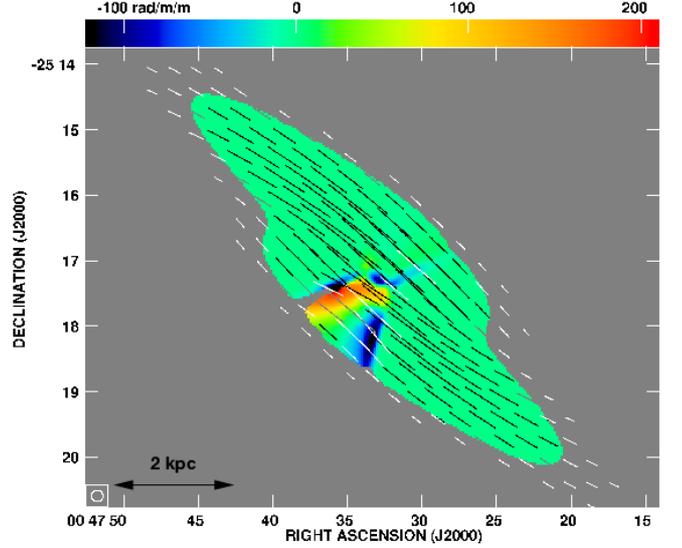}} \caption{Model RM
distribution and magnetic field orientation at $\lambda$3\,cm
at $10\farcs 5$ resolution as explained in the text.} 
\label{fig:rm_model}
\end{figure}

In this section we test whether the magnetic field in the outflow cone can
explain the RM enhancements around the minor axis seen in
Fig.~\ref{fig:radius}. No polarized emission from an extended halo field is
detected with our observations as the magnetic field vectors are mostly
parallel to the disc. The magnetic field only opens up farther away from the
galactic midplane and at larger galactocentric radii (see Paper~II).  However,
our high-resolution image (Fig.~\ref{fig:cm20tpCO12}) suggests that the
outflow cone has a strong magnetic field aligned with the walls of the cone,
so can influence RM, particularly in the SE cone that is located in front of
the bright polarized disc emission and may act as a Faraday screen. We made
models using an anisotropic magnetic field in the disc and a filamentary
magnetic field distributed in the walls of the conical outflow.  In
Table\,\ref{tab:cone_parameters} we summarize the parameters of the model, and
we show a sketch in Fig.\,\ref{fig:fil}. We assumed a rather large cone
opening angle of $45\degr$ as the cone should open farther away from the disc.
The model RM distribution is shown in Fig.\,\ref{fig:rm_model} along with the
magnetic field orientation. The magnetic field orientation is influenced
little by the magnetic field in the outflow cone because its $\rm |RM|$ of
$\le200\,\rm rad\,m^{-2}$ corresponds to a rotation of less than $46\degr$. In
the NW, the outflow cone is almost not detected in RM as it is located behind
the bright polarized disc. The vertical component of the magnetic field points
away from the disc in the SE outflow cone (see Paper~II).  However, a vertical
field alone cannot give rise to the observed strong change in RM. We also need
an azimuthal component parallel to the rotation direction of the disc. Both
components together form a \emph{helical} magnetic field, which generates the
strongest RM signals in the walls of the outflow cone. We assumed that the
azimuthal component rises linearly to a height of 1200\,pc, where the
azimuthal component is equal to the vertical field component (opening angle
$\alpha=45\degr$, see Appendix). As shown in Fig.\,\ref{fig:fil}, the field
lines are vertical close to the disc but start to wind up with increasing
height.
Our model can basically explain the observed RM distribution, especially the
RM jump between $60\degr$ and $80\degr$ in the middle ring ($1.0-2.2\,\rm
kpc$).  The red lines from the model shown in Fig.\,\ref{fig:radius} have the
right properties, but are not good fits. In the inner ring, the RM peak is at an
too small azimuthal angle, probably because the cone is not oriented perfectly
perpendicular to the disc. The RM amplitudes of the model are too small in the
middle ring, which indicates that the azimuthal field components increases
faster than linearly with radius. An improved model would need more parameters
than those given in Table\,\ref{tab:cone_parameters}, but with the present
observations we did not attempt to constrain them.
%
%
%
\subsection{Magnetic field strengths}
We estimated the magnetic field strength from the radio intensities by assuming
equipartition between the energy densities of the magnetic field and the total
cosmic rays \citep{beck_05a}. As we are only measuring the electrons we have
to make an assumption about the proton to electron ratio, $K=100$, the
usually assumed value for galaxies.
This choice of the $K$-factor is motivated by theory and
observations. Firstly, if the acceleration process produces a power law in
momentum of the cosmic rays and the initial numbers of protons and electrons
are equal, then one can infer that for relativistic energies the $K$-factor
depends only on the mass ratio of protons to electrons, as shown e.g., by
\citet[][Sect. 19.4]{schlickeiser_02a} and \citet[][their
appendix]{beck_05a}. Secondly, local measurements in the solar system can
directly lead to the proton and electron spectrum and corroborate
$K=100$. \citet{adriani_11a} present an electron spectrum based on the
satellite experiments PAMELA. From their Fig.\,1 one can infer the electron
flux as $0.18\,\rm s^{-1}\,m^{-2}\,GeV^{-1}$ (converting their
units). \citet{shikaze_07a} present a proton spectrum based on the balloon
experiment BESS. From their Fig.\,7, a proton flux of $20\,\rm
s^{-1}\,m^{-2}\,GeV^{-1}$ is inferred. Thus, the solar system value agrees
with $K=100$.
The $K$-factor is nevertheless somewhat uncertain for external galaxies and
thus induces uncertainties in the equipartition magnetic field
strength. Particularly, $K$ may be higher in case of strong energy losses of
the electrons, where the electrons have died away and only the protons (and
nuclei) remain. This may be the case for galactic haloes which contain an aged
population of cosmic-ray electrons. As we have seen in
Sect.\,\ref{subsec:spix}, the synchrotron losses in the nuclear outflow of
NGC\,253 are important, so that the $K$-factor may well be larger than
100; however the magnetic field strength depends on $K^{1/4}$, so
that a 50\% error in $K$ produces only a $13\%$ error in the magnetic field
strength. Summarizing, our magnetic field strengths should be taken as lower
limits.
The radio intensities were taken at $\lambda\lambda$ 20, 6, and 3\,cm
from the fits to the filaments, as shown in Fig.\,\ref{fig:slices}, and from the
maps in Figs.\,\ref{fig:20cm_tp}-\ref{fig:3cm_tpa}, where we subtracted the
background emission from the disc. The thermal contribution that also needs
to be subtracted (expect for the nucleus) was assumed to be 10\% ($\lambda
20\,\rm cm$), 20\% ($\lambda 6\,\rm cm$), and 30\% ($\lambda 3\,\rm cm$).  The
nuclear region has a size of $600\,\rm pc$, which we took as the integration
length along the line-of-sight, resulting in a magnetic field strength of
$160\pm 20\,\mu\rm G$. For the region of the outflow cone we took $500\,\rm
pc$, which is the diameter of the outflow cone. The result is $B_{\rm
  tot}=46\pm10\,\mu\rm G$ for the total field and $B_{\rm ord}=21\pm 5\,\mu\rm
G$ for the ordered field. The intensity of the filaments F1 and F2, observed
at $\lambda$20\,cm at high resolution (Fig.\,\ref{fig:cm20tpCO12}) with a
width of $< 40$\,pc, yields a lower limit of the total field strength of
$40\,\mu\rm G$ and $18\,\mu\rm G$ for the ordered field.
\begin{table}[tbhp]
\caption{Equipartition magnetic field strengths in various regions.}
\begin{center}
\begin{tabular}{lrrr}
\hline\hline
Region\T & $B_{\rm tot}$ & $B_{\rm ord}$ & Pol.\ degree\\
& [$\mu\rm G$] \B & [$\mu\rm G$] & [$\%$]\\\hline
Nuclear region \T & $160\pm 20$ & $<5$ & $<0.2$\\
Filaments F1+F2 & $46\pm 10$ & $21\pm 5$ & $21\pm 7$\\
Eastern bar & $24\pm 5$ & $11\pm3$ & $22\pm 7$\\
Western bar & $28\pm 6$& $5\pm 2$ & $4\pm 1$\\
\hline
\end{tabular}
\end{center}
\small{Note: the polarization degree of the synchrotron emission
refers to $\lambda$3\,cm, assuming 30\% thermal contribution}
\label{tab:equipartiton}
\end{table}
\section{Discussion}
\label{sec:discussion}
\subsection{Cosmic-ray transport}
\label{subsec:cr_transport}
As found in Sect.\,\ref{subsec:rm_cone}, the helical magnetic field lines are
confined to the walls that border the outflow cones, which appear as filaments
when observed in projection. Therefore, we can study the cosmic-ray transport
both, parallel to the filament direction and also perpendicular to it.
Galactic cosmic-ray electrons with energies of a few GeV lose
their energy mainly by synchrotron and inverse Compton (IC)
radiation, and by adiabatic losses. Other losses like ionization
ones can be neglected for the typical conditions of the
interstellar medium. The conditions inside the nuclear region are
different, and ionization losses may dominate
\citep{thompson_06a}, but we only investigate relativistic
electrons here that have escaped from the nuclear region.
The combined synchrotron and IC losses for a single cosmic-ray
electron is given by
\begin{equation}
\frac{{\rm d}E}{{\rm d}t} = - \frac{4}{3} \sigma_{\rm T} c \left ( \frac{E}{m_{\rm e}c^2}
\right )^2 (U_{\rm ph} + U_{\rm B}),
\end{equation}
where $U_{\rm ph}$ is the photon energy density, $U_{\rm B}$ the magnetic
field energy density, $\sigma_{\rm T} = 6.65 \times 10^{-25}\,\rm cm^2$ is the
Thomson cross section, and $m_{\rm e} = 511\,\rm keV\,c^{-2}$ the electron
rest mass. The time dependence of the energy is $E(t) = E_0 /(1+t/t_{\rm e})$,
where the electron lifetime is defined as the time at which the electron has
lost half its initial energy
\begin{equation}
  t_{\rm e} = 5.0 \times 10^{-4} \cdot \left ( \frac{\rm GeV}{E_0}\right ) \cdot
  \left (\frac{\rm erg\,cm^{-3}}{U_{\rm ph} + U_{\rm B}}\right ) \,\rm yr.
\end{equation}
Assuming that the cosmic-ray electrons emit most of their energy in
synchrotron emission at the critical frequency, the electron energy can be
expressed by:
\begin{equation}
E = \left ( \frac{\nu \cdot B}{{(16.1\,\rm MHz)} {(\mu\rm G)}}\right )^{1/2} \,\rm GeV.
\end{equation}
To calculate the electron lifetime in the filaments we use $U_{\rm
  B}=1.0\times 10^{-10}\,\rm erg\,cm^{-3}$ for $B=50\,\mu\rm G$ and $U_{\rm
  ph}=4.6\times10^{-11}\,\rm erg\,cm^{-3}$. We estimated the latter from the IRAS
FIR flux densities of the nucleus at a height of 300\,pc using cylindrical
symmetry. The IC and synchrotron losses are therefore
comparable. The electron lifetimes are 2.5 ($\lambda$20\,cm), 1.4
($\lambda$6\,cm), and $1.1\times 10^6\,\rm yr$ ($\lambda$3\,cm).
For dominating radiation losses, the cosmic-ray electron
scaleheight is a function of the frequency
\begin{eqnarray}
  h_{\rm e} & \propto & \nu^{-(1-a)/4}\quad {\rm (diffusion)}\nonumber\\
  h_{\rm e} & \propto & \nu^{-0.5} \quad {\rm (convection)},
\end{eqnarray}
depending on whether we have diffusion or convection as the transport
mechanism. These scaleheights are derived as follows. We note that in the
equipartition case we have $h_{\rm e}=2h_{\rm syn}$ (Eq.\,11 in Paper~I). For
diffusion, $a$ is the energy dependency of the diffusion coefficient, usually
assumed to be $a\approx 0.5$ for Kolgomorov type turbulence
\citep{schlickeiser_02a}. For the case of convection, the scaleheight is
proportional to the electron lifetime with $t_{\rm e} \propto \nu^{-1/2}$. In
the case of diffusion we have $h^2_{\rm e}/t_{\rm e}=\kappa$, and with $\kappa
\propto E^a$ and $t_{\rm e}\propto 1/E$ we get the above relation.
The observed frequency dependence of the scaleheight in the filaments is
$h_{\rm e}\propto \nu^{0\pm 0.16}$, which is not consistent with convection in
a galactic wind. Diffusion along the filaments would require $a=1\pm 0.64$,
which can not be excluded a priori. However, diffusion should result in a
Gaussian intensity profile, while the observed one is clearly exponential.
Furthermore, the filament width of $\approx 300\,\rm pc$
(Sect.\,\ref{subsec:emission_profiles}) is larger than its scaleheight. If
cosmic-ray transport is responsible for the shape of the filaments, we would
expect that the filament scaleheight is greater than the filament width,
because diffusion along the field lines is much faster than perpendicular to
them. We conclude that the cosmic-ray electrons do not propagate along the
filaments, if radiation losses dominate. We give an alternative explanation in
Sect.\,\ref{subsec:radio-fir}.
The filaments have a Gaussian profile, as expected for diffusion, and their
widths were determined in Sect.\,\ref{subsec:emission_profiles} as a function
of frequency as $\propto \nu^{-0.13\pm 0.18}$. This is consistent with
diffusion {\em perpendicular}\ to the filaments with $a = 0.5\pm 0.7$, close
to the expected behaviour, albeit with a large error interval. It seems
therefore straightforward to assume that the cosmic-ray electrons of the
surrounding medium diffuse from outside into the filaments. This explains why
the filaments have a scaleheight nearly independent of the frequency, but their
width decreases with increasing frequency (but see also
  Sect.\,\ref{subsec:radio-fir}). Summarizing, we can explain the
observations by \emph{magnetic filaments} with a width of about $40\,\rm pc$
(or less) at the boundary of both outflow cones. The perpendicular diffusion
coefficient across the filaments is $\kappa_{\rm \perp} = 1.5\times
10^{28}\,{\rm cm^2\, s^{-1}} \cdot E({\rm GeV})^{0.5\pm0.7}$.
We can compare this result with theoretical expectations as the recent
paper by \citet{shalchi_10a}, who used numerical calculations to get an
estimate for the perpendicular diffusion coefficient in a turbulent
interstellar medium. Our cosmic-ray electron energy is a few GeV, so that for
protons the magnetic rigidity $R=(p\cdot c)/(Z\cdot e)$ is $R\approx 5\times
10^{12}\,\rm Volt$. For this rigidity \citet{shalchi_10a} give a perpendicular
diffusion coefficient of $\kappa_{\rm \perp} = 3\times 10^{27}\,{\rm cm^2\,
  s^{-1}}$, a factor of 10 smaller than what we find. However, the
theoretically expected parallel diffusion coefficient is a factor of 100
larger. A small amount of turbulence in the filaments is able to explain our
observations.
\subsection{Energetics in the nuclear outflow}
The nuclear outflow in NGC\,253 appears to be a powerful version of the
outflow in our own Milky Way. The supernova rate in NGC\,253 lies between 0.03
and 0.3\,$\rm yr^{-1}$ \citep{ulvestad_97a,lenc_06a} and is thus much larger
than for the Galactic centre with $(0.02-0.08)\times 10^{-2}\,\rm yr^{-1}$
\citep{crocker_11a}. Assuming that the outflow is driven by the starburst, the
energy input in NGC\,253 is thus a factor of 100 higher than in the Milky
Way. Nevertheless, the Milky Way is able to launch a powerful galactic wind
from its central region, as indicated by the strong deficiency in the radio
emission compared to the FIR luminosity \citep{crocker_11b}. Also,
the gamma ray emission is deficient compared to the star formation and
supernova rates, which can be explained by the advection of cosmic rays in the
wind, escaping into the halo.
\citet{dobler_08a} have detected a microwave haze in WMAP observations towards
the Galactic centre, which they interpreted as non-thermal synchrotron emission
from cosmic-ray electrons. \emph{FERMI}-LAT observations have revealed two
giant gamma-ray bubbles extending to a Galactic latitude of $50\degr$, and
containing an energy of $10^{54}-10^{55}\,\rm erg$
\citep{dobler_10a,su_10a}. These discoveries were preceded by the finding of a
bipolar outflow in X-ray emission by \citet{bland-hawthorn_03a}, estimating a
similar amount of energy released. To create these bubbles, the current
release of energy by star formation is not enough. However, the bubbles have
an age of $10^7\,\rm yr$, so that a starburst in the last $10\,\rm Myr$ could
explain them. The alternative possibilities are past accretion events of tidal
debris of disrupted stars by the central black hole, which can also explain
the presence of hot X-ray emitting gas \citep{totani_06a}.
In the Galactic centre there are lobes of radio continuum emission with a
diameter of 110\,pc and a height of 165\,pc \citep{law_10a}. The Galactic
centre lobes look very similar to the structure we observe in NGC\,253, as can
be seen by comparing Fig.\,1 in \citet{law_10a} with our
Fig.\,\ref{fig:20cm_xray}, although the filaments in NGC\,253 are a factor of
2 larger. The equipartition field strengths between 40 and 100\,$\mu$G are
similar between the two galaxies. We note that we can trace the extension of
the filaments in NGC\,253 to a much larger height of 600\,pc
(Fig.\,\ref{fig:slices}), likely because our beam size of 150\,pc is much
larger than that of \citet{law_10a}. Observations of the Faraday rotation
favour a strong vertical magnetic field in the Galactic centre lobe region
\citep{law_11a}.
We conclude that the radio continuum morphology of the Galactic centre and the
nuclear starburst in NGC\,253 are similar, despite their current difference in
star formation rate. As the nuclear outflow in NGC\,253 can be explained by
the starburst model alone, the similarity may hint that the Milky Way bipolar
outflow seen in X-ray and gamma-ray emission is also generated by a past
starburst, rather than the past activity of the Galactic central black hole. Gamma
rays are now also detected from the nucleus in NGC\,253 by H.E.S.S., a
\v{C}erenkov air shower telescope \citep{acero_09a}. According to their
results, the nucleus is not a good ``calorimeter'', indicating a loss of
cosmic rays, possibly enlightening a bipolar outflow seen in gamma rays as in
the Milky Way.
\begin{figure*}[tbhp]
\resizebox{\hsize}{!}{ \includegraphics{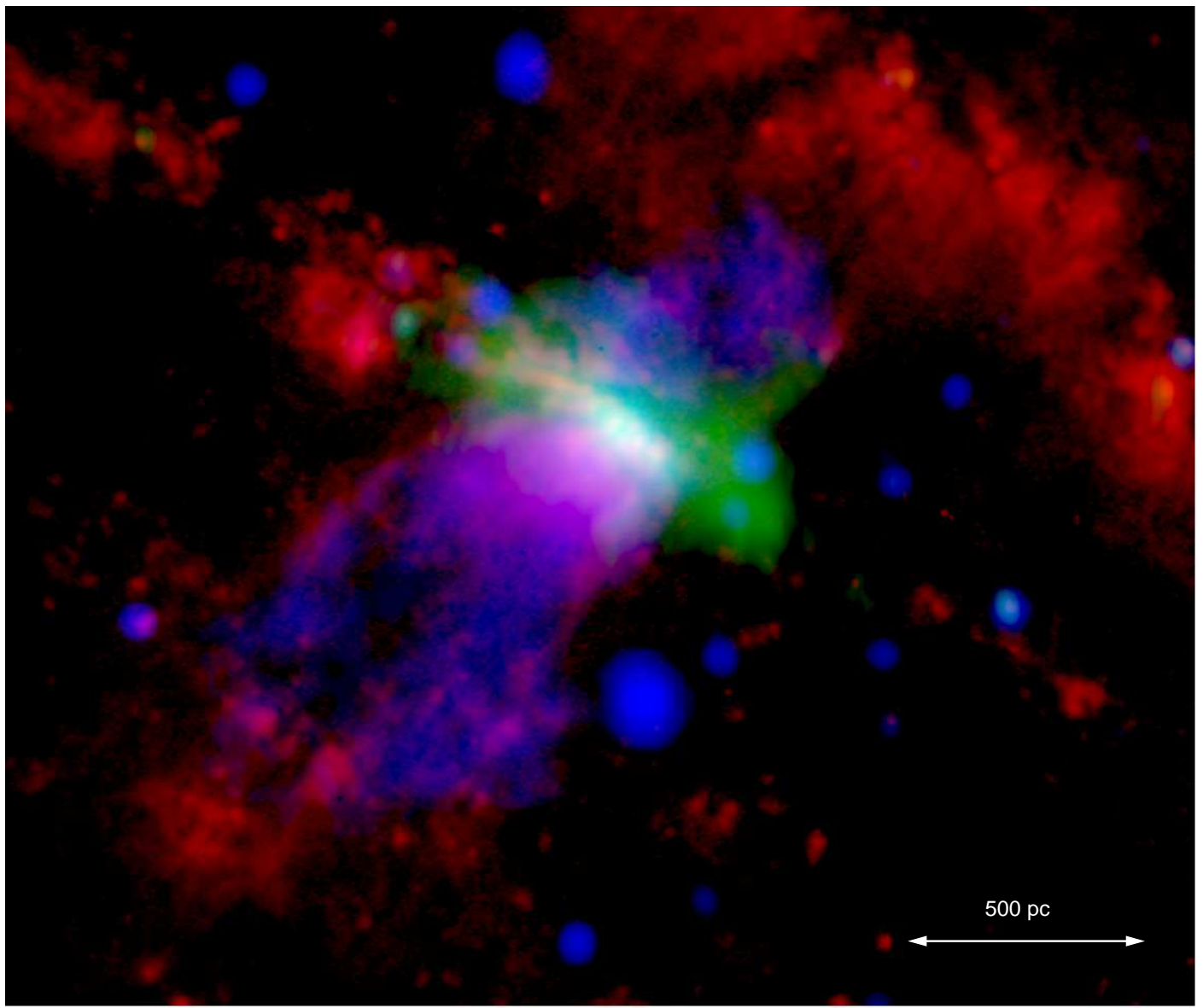}}
\caption{Three-colour composite, multi-wavelength view of the central region in
  NGC\,253. Red, green, and blue indicate H$\alpha$ from \citet{westmoquette_11a},
$\lambda$20\,cm radio continuum, and \emph{Chandra} soft X-ray
(M. Hardcastle, priv.\ com.), respectively.}
\label{fig:rgb}
\end{figure*}
\subsection{Radio-FIR correlation}
\label{subsec:radio-fir}
As derived from IRAS flux densities, the nucleus of NGC\,253 has a FIR
luminosity of $FIR/(3.75\times 10^{12}\,\rm Hz)=4.9\times 10^{-8}\,\rm
erg\,s^{-1}\,cm^{-2}$. Converting this into the radio (spectral) luminosity at
1.4\,GHz, using the radio-FIR correlation \citep[e.g.,][]{condon_02a}, we find
$S_{\rm 1.4\,GHz}=6.5\,\rm Jy$. Our observed radio luminosity of the nucleus
of $2\,\rm Jy$ (Paper~I) thus falls a factor of 3 short of the
expectation. This implies that the nucleus is not a ``calorimeter'' in the radio
regime \citep{voelk_89a}, and that non-synchrotron losses of the cosmic-ray
electrons could be important.
Such a process, which we expect to be strong in a galactic wind, is
adiabatic cooling in an accelerated flow. If cosmic rays are
advectively transported, the adiabatic loss time is
\begin{equation}
t_{\rm ad} = 3 \left ( \frac{\rm dv}{{\rm d}z} \right )^{-1},
\end{equation}
where ${\rm dv}/{\rm d}z$ is the velocity gradient in the vertical
direction. We note that the adiabatic loss time does not depend on the
observing frequency, in contrast to synchrotron and IC losses. If
adiabatic losses are dominating, we could naturally explain the independence
of the synchrotron scaleheight from the observing frequency
(Sect.\,\ref{subsec:emission_profiles}). As an upper limit for $t_{\rm ad}$ we
take the cosmic-ray electron lifetime at $\lambda$3\,cm of 1\,Myr, as derived
in Sect.\,\ref{subsec:cr_transport}. We use a cosmic-ray bulk speed of
$300\,\rm km\,s^{-1}$ as derived from the halo scaleheights (Paper~I),
which was confirmed by \citet{zirakashvili_06a} for the nuclear
outflow. Assuming linear acceleration (with respect to $z$), we find that the
cosmic rays are accelerated between 0 and 100\,pc height.
We can also estimate the electron scaleheight as $h_{\rm e}= {\rm v}\cdot
t_{\rm ad}$ in the advection flow. The derived value of $h_{\rm e}=300\,\rm
pc$ fits to the synchrotron scaleheight of 150\,pc, if the magnetic field
scaleheight is much larger, such as the 600\,pc expected for energy
equipartition. Since our estimated electron scaleheight fits the observations,
the electron lifetime of 1\,Myr is realistic. In this case, the adiabatic
losses would be comparable to the combined synchrotron and IC losses. As the
last two have about the same magnitude (Sect.\,\ref{subsec:cr_transport}),
we can explain a radio deficiency of a factor of 3. We conclude
that adiabatic losses in the nuclear outflow can explain the dim radio
luminosity of the nucleus. But other processes may still contribute, such as
ionization and bremsstrahlung losses of the cosmic rays in the nucleus
\citep{lacki_11a}.
\subsection{Collimating the nuclear outflow}
The SE outflow cone has an opening angle of $26\degr$ as H$\alpha$ and X-ray
data show \citep{strickland_00a,bauer_07a}. The X-ray data from the NW cone
suggest a similarly small value although with a larger
uncertainty. Interaction with the surrounding interstellar medium from the
galactic disc can explain such a small cone opening
angle. Figures \ref{fig:3cm_tpa} and \ref{fig:20cm_xray} show that the
filaments of the radio emission border the outflow cones visible in the hot
X-ray emitting gas. The H$\alpha$ emitting gas borders again the X-ray
outflow. In Fig.\,\ref{fig:rgb} we present a multi-wavelength composite of the
central region of NGC\,253, which nicely shows the interplay between the
different components of the interstellar medium. The radio filaments clearly
surround the outflow cones at least near the base. Therefore, it is useful
to ask whether the magnetic field can collimate the outflow cone?
To answer this question we
make an estimate of the energy densities.  \citet{bauer_07a} obtained an
electron density of $n_{\rm e}=0.025\,\rm cm^{-3}$ at a temperature of
$0.38\,\rm keV$ in the SE cone, resulting in an energy density of $1.5\times
10^{-11} \rm erg\, cm^{-3}$. The NW cone is brighter with a temperature of
$0.54\,\rm keV$ so that here the energy density is $3.0\times 10^{-11} \rm
erg\, cm^{-3}$ (assuming also an increased electron density). The minimum
estimate of the magnetic field strength in the filaments is $50\,\mu\rm G$
resulting in an energy density of $1.0\times 10^{-10} \rm erg\, cm^{-3}$. This
is higher than the thermal energy density, so that the magnetic field can
explain the collimation of both outflow cones.

\subsection{Galactic dynamo}
The parity of the magnetic fields can be used to test models of the galactic
dynamo. The expectation is that a mean-field dynamo operating in a thin disc
gives rise to a quadrupolar, even-parity magnetic field, where the direction
of the azimuthal magnetic field is the same above and below the plane, but the
direction of the vertical magnetic field component reverses with respect to
the plane \citep[e.g.,][]{krause_89b,haverkorn_11a}. This also applies to a
thin disc embedded in a quasi-spherical halo \citep{moss_92a}. However, a
dynamo operating in a quasi-spherical halo should give dipolar, odd-parity
magnetic fields, with an azimuthal magnetic field with reversing direction
across the Galactic plane, while the vertical field is directed in the same
sense above and below the plane \citep{sokoloff_90a}. So far the observational
evidence prefers even field parities in galaxies, based on the RM patterns in
the disc \citep[e.g.,][]{krause_89a,krause_89b,tabatabaei_08a} or the
asymmetry of polarized emission \citep{braun_10a}. These RM measurements of
moderately inclined galaxies are, however, no stringent clue, because the RM
structures of an even and odd parity in the disc fields are very similar in
our synthetic polarization and RM maps (Figs.\,A.1 and A.2 in Paper~II).

RM measurements in halo magnetic fields, which allow better distinction, are
more difficult due to sensitivity and overlap with the disc field. Such
measurements do also favour quadrupolar, even-parity fields
\citep{soida_11a}. In Paper~II we argued that NGC\,253 is another candidate of
an even disc field, albeit limited by the low resolution. This paper gives
much better evidence for an even field in the disc beyond about 2\,kpc
radius. High spatial resolution is crucial to distinguish between even and odd
parity of the disc magnetic field. The synchrotron scaleheight of the thin
disc is about $300\,\rm pc$ (Paper~I), so that we need a comparable
resolution, like $6\arcsec$ ($D=10\,\rm Mpc$), to resolve an RM reversal
across the midplane.  Such high angular resolutions {\em with sufficient S/N}
in polarization now become feasible with the advent of new broadband
correlators at various radio interferometers. We expect that future
observations will be better able to detect odd disc fields in galaxies, some of
which may have so far been classified as even.
We detected anisotropic magnetic fields as the dominant contribution to the
ordered disc field observed in polarized intensity within a radius of $\approx
2\,\rm kpc$.  Anisotropic magnetic fields can be generated from random,
isotropic fields by shear and compression. M\,31 is a galaxy that is dominated
by a coherent axisymmetric magnetic field field from its RM structure
\citep{berkhuijsen_03a}. In contrast, M\,51, a grand-spiral galaxy with a
prominent spiral magnetic field, shows almost no large-scale RM pattern
\citep{fletcher_11a}. This galaxy is a prime example of an anisotropic disc
magnetic field. The late-type spiral galaxy NGC\,6946 has both regular
magnetic fields between the spiral arms, in the so-called ``magnetic arms'',
and anisotropic magnetic fields \citep{beck_07a}. Barred galaxies, such as
NGC\,1097 and NGC\,1365, with their strong shearing flows host dominant
anisotropic fields \citep{beck_05b}. In light of these results, it comes as no
surprise that the ordered disc field in NGC\,253 is anisotropic within
$\approx 2\,\rm kpc$ radius, where its bar resides (Sect.\,\ref{subsec:bar}).
\subsection{Helical magnetic field in the outflow cone}
The helical field can be explained with help of trajectories of particles in
the outflow. The conductivity of the interstellar medium is very high, so that
the magnetic field lines are essentially frozen into the plasma.  The field
lines are taken with the gas in the outflow similar to the solar wind
\citep{parker_58a,weber_67a}. The magnetic field is anchored in the disc
between a radius of about $130-170$\,pc (Fig.\,\ref{fig:fil}). The cone
opening angle of $26\degr$ ($45\degr$ at larger heights) means that the field
is transported to larger radii farther away from the disc. If we assume that
no significant interaction with the surrounding medium occurs, the angular
momentum of the plasma in the outflow is conserved. From the perspective of an
outflowing particle, the anchor point rotates faster than the particle. The
trajectories of the particles can be described by a spiral, similar to water
drops in a garden sprinkler. The spiral orientation is given by the sense of
rotation and is identical to that of the spiral arms, i.e.\ trailing arms. The
vertical component of the magnetic field has to point away from the disc in
order to explain the observed RM structure, which agrees with our
results in Paper~II for the SE halo.
We can compare the actual shape of our helix with this model. The rotation
curve of the galaxy is rising linearly out to a radius of 300\,pc, where it
tailors off to a rotation speed of about $200\,\rm km\,s^{-1}$
\citep{sorai_00a}. The rotation speed at the anchor point of the filaments is
thus $\approx 100\,\rm km\,s^{-1}$, resulting in a rotation period of
$0.9\times 10^7\,\rm yr$. We measured that the azimuthal component is about
equal to the vertical component at a height of 1200\,pc
(Sect.\,\ref{subsec:rm_cone}). The circumference at 150\,pc radius is 940\,pc,
similar to the height, so that we can expect an equal strength of the vertical
and azimuthal components. During one rotation period, the particles in the
outflow should thus be transported to a height of 1200\,pc. The outflow speed
in the nuclear outflow is $\approx 300\,\rm km\,s^{-1}$
\citep{westmoquette_11a} at a reference height of 1\,kpc, so that assuming
constant acceleration the average speed is $150\,\rm km\,s^{-1}$. To reach a
height of 1200\,pc a particle would take a time of $0.8\times 10^7\,\rm yr$,
which is in good agreement with the rotation period.
Our simple model of the magnetic field lines frozen into the outflow plasma
can therefore explain the observed winding-up of the field lines into a
helix. We also note that the Alfv\'en speed ${\rm v_{\rm A}} = B /
\sqrt{4\pi\rho}$ is $\rm 30\,\rm km\,s^{-1}$ with $n_{\rm e}=2\,\rm cm^{-3}$
and $B=21\,\rm\mu G$. This is only 10\% of the outflow speed at a reference
height of 1\,kpc, so that the Alfv\'enic point would be with 100\,pc close to
the disc. The stiff field lines co-rotate with the disc within the Alfv\'enic
point. This means that the field lines start almost immediately winding up in
the outflow, which accords with our observations.
Although the nuclear outflow in NGC\,253 is thought to be driven entirely
by a starburst and not by an active galactic nucleus
\citep[AGN,][]{brunthaler_09a}, it is worthwhile briefly comparing the
outflow structure with those of AGNs. There are a number of radio
polarimetric studies of AGNs, employing very long baseline interferometry and
studying their magnetic fields using the synchrotron emission of highly
energetic electrons in the jet \citep[e.g.,][]{gabuzda_92a,gabuzda_00a}. They
show that the magnetic field orientations are mainly transversal to the jet
direction. This is thought to be caused by compression in shock waves, where
the cosmic-ray electrons are accelerated, and thus illuminate the ``knots''
in the jets where the shocks occur. However, the alternative explanation is
that the transverse fields are generated by the differential rotation in the
accretion disc around the black hole \citep[e.g.,][]{contopoulos_09a}. This would
result in a helical magnetic field farther away from the accretion disc.
One signature of a helical magnetic field in the jet of AGNs should be a reversal
of the RM across the jet, as observed by us in the nuclear outflow of
NGC\,253. Studies of the RM in jets of AGNs are notoriously difficult, given
that the jets are mostly only barely resolved and the thermal electron
densities are low \citep{gabuzda_04a,kharb_09a,sullivan_09a}. However, these
authors find a strict dependence of the polarization angle on $\lambda^2$,
suggesting that the jets form a Faraday screen and that the observed RM
gradients are genuine and not instrumental. If the view onto the jet is nearly
``side-on'' ($90\degr$ to its axis) in its rest frame, the magnetic field
would be transversal. If the jet is observed at a different angle to
$90\degr$, we should see a central region with a transversal magnetic field,
whereas the field should be along the jet direction at the boundary. Such
``spine + sheat'' structures have been indeed observed
\citep[e.g.,][]{attridge_99a,pushkarev_05a}, further supporting the existence
of helical fields in AGN jets.
\section{Conclusions}
\label{sec:conclusions}
We used polarimetric radio continuum observations to study the nuclear outflow
in NGC\,253. Four prominent filaments were detected, which we interpreted as
  the walls of the two nuclear outflow cones seen in projection. The
  boundary of the NW outflow cone is not seen in H$\alpha$ and X-ray emission,
  probably due to heavy absorption. These are our main conclusions
\begin{enumerate}
\item The filaments have a scaleheight of $150\pm 20\,\rm pc$, independent of
  wavelength. The cosmic-ray electron lifetime changes by a factor of $2.3$
  between $\lambda\lambda$ 20 and 3\,cm, so that the scaleheight should change
  if the electrons are subjected to dominant radiation losses. Dominating
  adiabatic losses could explain this behaviour, where cosmic rays and
  magnetic fields are advected together in the accelerating nuclear outflow.
\item The filaments have a smaller perpendicular extent for smaller electron
  lifetimes, so that perpendicular electron diffusion plays a role. This
  suggests that we are looking at \emph{filamentary} magnetic fields. The
  perpendicular diffusion coefficient is $\kappa_{\rm \perp} = 1.5\times
  10^{28}\,{\rm cm^2\, s^{-1}} \cdot E({\rm GeV})^{0.5\pm0.7}$.
\item The filaments have a total magnetic field strength of $B_{\rm
    tot}=46\pm10\,\mu\rm G$ and ordered field strength of $B_{\rm ord}=21\pm
  5\,\mu\rm G$. The field strengths could be higher,
  if the cosmic rays are not in equipartition with the magnetic field.
\item The magnetic field strength in the filaments is high enough to
  counteract the pressure in the nuclear outflow. The magnetic field can thus
  collimate the outflow and explain the small opening angle of $\approx
  26\degr$ by interaction with the interstellar medium. The magnetic field is
  aligned along the filaments, as expected for the case of compression due to
  a lateral pressure gradient.
\item In the SE outflow cone, the magnetic field points away from the disc
  in the form of a helix. The magnetic field is frozen into the plasma of
    the nuclear outflow, so that the angular momentum is conserved and the field
    wound up.
  \item The ordered magnetic field in the disc observed in polarized intensity
    is anisotropic within a radius of about 2\,kpc. At larger radii a regular
    field with an even axisymmetric pattern is found, as already deduced in
    Paper~II.
  \end{enumerate}
\begin{acknowledgements}
  VH is funded as a postdoctoral research assistant by the \emph{Science and
    Technology Facilities Council (STFC)} under a rolling grant. RJD is
  supported by the \emph{Deutsche Forschungsgemeinschaft (DFG)} in the
  framework of the research unit FOR\,1048. This work benefits from various
  ancillary data provided kindly by colleagues We thank Kazushi Sakamoto for
  providing us the CO map, Charles Hoppes and Mark Westmoquette for the
  H$\alpha$ maps, and Martin Hardcastle for the \emph{Chandra} X-ray map. We
  thank Uli Klein for corrections to the manuscript at an early stage. We are
  grateful to Elias Brinks for carefully reading the manuscript. Wolfgang
  Reich is thanked for many useful suggestions that helped to improve the
  paper.
\end{acknowledgements}

\bibliography{vh3_bib}

This paper has been typeset from a \TeX/\LaTeX file prepared by the author.

\appendix

\section{Synthetic polarization maps}
\label{app:synthetic_polarization_maps}
In \citet{heesen_09b} we used a conical halo field with a filled cone while in this
paper we use a conical field that is only different from zero inside the walls
of the outflow as shown in Fig.\,\ref{fig:fil}. The following equations are therefore
identical for both cases. We stress that, even though we denote the vertical
magnetic field component with the superscript ``halo'', this field is not a
general large-scale halo field but is only concentrated in the filaments. We do
not use any turbulent magnetic field, but instead restrict ourselves to ordered
magnetic fields.
We integrate the source structure in a three-dimensional volume with Cartesian
coordinates $(x,y,z)$, where the $x$-axis is parallel to R.A., $y$ is parallel to the
line-of-sight, and $z$ parallel to Dec. We also define the coordinates
$(x',y',z')$ that are the Cartesian coordinates with respect to the galaxy. $x'$
is parallel to the major axis, $y'$ is parallel to the minor axis, and $z'$
parallel to the rotation axis. We can relate both coordinate systems by
linear transformations first rotating around the inclination angle $i$ and
then rotating it around the position angle. For convenience we use $\iota
=90\degr-i$, which is the angle between $y$ and $y'$, and $\zeta =90\degr - p.a.$
which is the angle between $z$ and $z'$. The reference coordinates $(x_{\rm
  ref},y_{\rm ref},z_{\rm ref})$ are the central coordinates of the
galaxy. Moreover, we use cylindrical coordinates for the observer
$(\rho,\phi,z)$ and in the galaxy $(\rho',\phi',z')$.
\begin{figure*}[tbhp]
\resizebox{\hsize}{!}{ \includegraphics{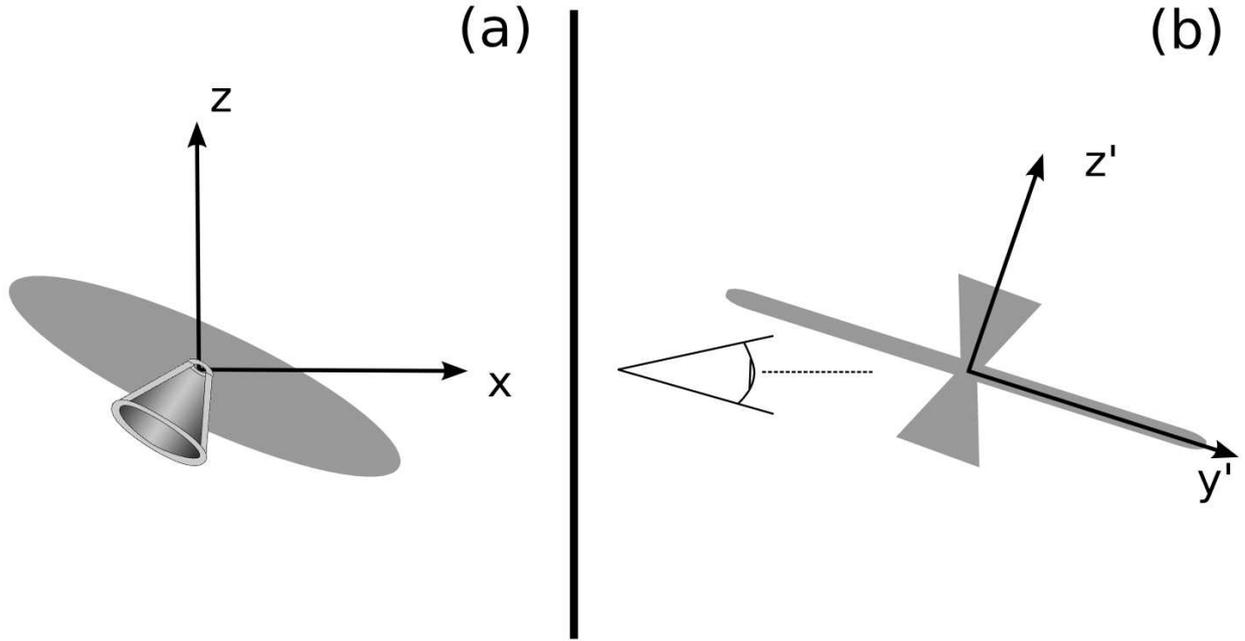}}
\caption{Coordinates used for the synthetic polarization maps. (a) View along
  the line-of-sight in the direction of the $y$-axis. (b) View from the side
  in the opposite direction to the $x'$-axis, where the observer is located at
  the left. }
\label{fig:coo}
\end{figure*}

The frame of reference Cartesian coordinates of the galaxy can be expressed in the observer's as
\begin{equation}
  x' = \cos(\zeta) \cdot (x-x_{\rm ref}) - \sin(\zeta) \cdot (z-z_{\rm ref})
\end{equation}
\begin{eqnarray}
  y' = -\sin( \zeta ) \cdot \sin( \iota ) \cdot ( x - x_{\rm ref} ) + \cos( \iota ) \cdot ( y -
  y_{\rm ref} )\nonumber \\
 - \sin( \iota ) \cdot \cos ( \zeta ) \cdot ( z - z_{\rm ref} )
\end{eqnarray}
\begin{eqnarray}
  z' = \sin( \zeta ) \cdot \cos( \iota ) \cdot ( x - x_{\rm ref} ) + \sin( \iota ) \cdot ( y -
  y_{\rm ref} ) +\nonumber \\
 \cos( \iota ) \cdot \cos( \zeta ) \cdot ( z - z_{\rm ref} ).
\end{eqnarray}
For the strength of the magnetic field in the disc we use a Gaussian profile with
respect to the galactocentric radius and an exponential vertical decrease with
a scaleheight $h_{\rm B, disc}$. For NGC\,253 we used a two-component Gaussian
distribution as a function of $\rho'$, which is not shown here for
simplicity. The parameter $\sigma$ can be expressed by the full width half
mean (FWHM) as $\sigma= {\rm FWHM}/(2\sqrt{2\log(2)})$ in this example and all
following cases where we use a Gaussian distribution:
\begin{eqnarray}
  B^{\rm disc} = B^{\rm disc}_0 \cdot \exp \left ( \frac{-|z'|}{h^{\rm disc}_{\rm B}} \right ) \cdot
  \exp\left ( \frac{-\rho'^2}{2 \sigma_{\rm disc}^2} \right ),
\end{eqnarray}
where we take the absolute value $|z'|$ of the vertical distance to
the galactic midplane. The magnetic field strength in the halo (filament)
can be expressed as the disc field by
\begin{eqnarray}
  B^{\rm halo} = B^{\rm halo}_0 \cdot \exp\left ( \frac{-|z'|}{h^{\rm halo}_{\rm B}} \right )
  \cdot \exp\left ( \frac{-\rho'^2} {2 \sigma_{\rm halo}^2} \right ).
\end{eqnarray}
Here, $B^{\rm disc}_0$ and $B^{\rm halo}_0$ are the strengths of the ordered
field in the disc and halo. The $z'$-component of the halo (filament) field
is
\begin{equation}
B'_{z, \rm halo} = \pm\frac{B'_{\rm halo}}{\sqrt{1.0 + \tan^2(\alpha) +
    \tan^2(\beta)}},
\end{equation}
where for even parity ``$+$'' is used if $z'\leq 0$ and for odd parity ``$+$''
is used for all $z'$. We can now derive the $\phi'$-component as
\begin{equation}
  B'_{\phi',\rm halo} = \pm B'_{z, \rm halo} \cdot \tan(\alpha),
\end{equation}
where ``$+$'' is used if $z' \leq 0$. The angle $\alpha$ gives the
azimuthal component of the halo (filament) field and $\beta$ is the angle with
respect to the rotation axis, i.e.\ half of the cone opening angle. Now
\begin{equation}
  B'_{\rho',\rm halo}  = \pm B'_{z, \rm halo} \cdot \tan(\beta),
\end{equation}
where ``$+$'' is used if $z' \leq 0$. Now
\begin{equation}
  B'_{\phi',\rm disc} = \pm \frac{B'_{\rm disc}}{1.0 + \tan^2(\chi)},
\end{equation}
where for even parity ``$-$'' is used for all $z'$ and for odd parity ``$+$''
is used if $z' \leq 0$. The azimuthal disc component is
\begin{equation}
  B_{\rho', \rm disc} = \pm \tan(\chi) \cdot B_{\phi',\rm disc},
\end{equation}
where $-$ is used for an inwards pointing field. For the following we define
$\varphi$ as
\begin{equation}
  \varphi = \rm{atan2}(y',x'),
\end{equation}
where the numerical function ``atan2'' is equivalent to atan, but returns the
angle in all four quadrants ($360\degr$) rather than only between $-90\degr$
and $90\degr$. Transforming back from the cylindrical coordinate system of the
galaxy into the Cartesian coordinate system of the observer we first need to
transform to the Cartesian system in the galaxy:
\begin{equation}
  (B^{\rm disc}_x)' = (B^{\rm disc}_\rho)' \cdot \cos( \varphi ) - (B^{\rm disc}_\phi)' \cdot
  \sin( \varphi )
\end{equation}
\begin{equation}
(B^{\rm disc}_y)' = (B^{\rm disc}_\rho)' \cdot \sin( \varphi ) + (B^{\rm disc}_\phi)' \cdot
\cos( \varphi ).
\end{equation}
Transforming above equations into the Cartesian system of the observer,
\begin{eqnarray}
  B^{\rm disc}_x & = & \cos(\zeta) \cdot (B^{\rm disc}_x)' - \sin(\zeta) \cdot
  \sin (\iota) \cdot (B^{\rm disc}_y)' + \nonumber\\
& & \sin(\zeta) \cdot \cos(\iota)
  \cdot (B^{\rm disc}_z)'
\end{eqnarray}
\begin{equation}
  B^{\rm disc}_y = \cos(\iota) \cdot  (B^{\rm disc}_y)' + \sin(\iota)  \cdot
  (B^{\rm disc}_z)'
\end{equation}
\begin{eqnarray}
  B^{\rm disc}_z & = & -\sin(\zeta) \cdot (B^{\rm disc}_x)' - \sin(\iota) \cdot
  \cos (\zeta) \cdot (B^{\rm disc}_y)' + \nonumber\\
& & \cos(\iota) \cdot \cos(\zeta) \cdot (B^{\rm disc}_z)'.
\end{eqnarray}
The same equation applies to the halo (filament) magnetic field that is not
shown here. Now we have the magnetic field components in the Cartesian system
of the observer. We define the local polarization angles for the disc and halo
field as
\begin{equation}
  \psi_{\rm disc} = {\rm atan2}\left \lbrace -(B^{\rm disc}_x )', (B^{\rm
      disc}_z)' \right \rbrace - 90\degr
\end{equation}
\begin{equation}
  \psi_{\rm halo} = {\rm atan2} \left \lbrace -(B^{'\rm halo}_x)', (B^{'\rm
    halo}_z)' \right \rbrace - 90\degr .
\end{equation}
The $90\degr$ correction is to transform the polarization angle into the magnetic field orientation. The Faraday rotation shifts the polarization angle if we go a step $\Delta y$ along the
line-of-sight:
\begin{eqnarray}
  \Delta \psi(y+\Delta y) & = & \Delta \psi(y) - 0.81 \cdot \lambda^2 \cdot n_{\rm e} \cdot
\exp\left ( \frac{-|z'|}{h_{\rm e}} \right ) \exp \left ( \frac{\rho'^2}
  {2\sigma_{\rm e}^2} \right ) \cdot \nonumber\\
& & \left \lbrace \left ( \frac{B_{\rm reg}}{B_{\rm ord}} \right )^{\rm
  disc} \cdot (B^{\rm disc}_y)' +\right . \nonumber\\
& & \left . \left ( \frac{B_{\rm reg}}{B_{\rm ord}} \right )^{\rm
  halo} \cdot (B^{\rm halo}_y)' \right
\rbrace\cdot \Delta y.
\end{eqnarray}
The ``$-$'' is because we are going in positive y-direction, whereas the
polarized wave is going from the source to the observer. The ratio of the
regular to the ordered field describes the anisotropy of the magnetic field,
with $B_{\rm reg} \leq B_{\rm ord}$.
We can now also calculate the Faraday screen RM if the galaxy would be a
screen:
\begin{eqnarray}
  RM_{\rm screen}(y+\Delta y) & = & RM_{\rm screen}(y) - 0.81 \cdot n_{\rm e} \cdot
\exp\left ( \frac{-|z'|}{h_{\rm e}} \right )\nonumber\\
& & \exp \left ( \frac{\rho'^2} {2\sigma_{\rm e}^2} \right ) \cdot \left
  \lbrace \left ( \frac{B_{\rm reg}}{B_{\rm ord}} \right )^{\rm
  disc} \cdot (B^{\rm disc}_y)' + \right . \nonumber\\
& & \left . \left ( \frac{B_{\rm reg}}{B_{\rm ord}} \right )^{\rm
  halo} \cdot (B^{\rm halo}_y)' \right \rbrace \cdot \Delta y.
\end{eqnarray}
For the observed polarization angle we obtain with Faraday rotation included:
\begin{equation}
  \psi^{\rm Faraday}_{\rm disc} = \psi_{\rm disc} + \Delta\psi .
\end{equation}
The cosmic-ray energy density $n_{\rm CR}$ is scaled according to the
equipartition condition as
\begin{eqnarray}
  n_{\rm CR} & = & n_{\rm CR,0} \left ( \frac{(B^{\rm disc}_x)^2+(B^{\rm disc}_y)^2+(B^{\rm
      disc}_z)^2}{(B^{\rm disc}_0)^2+(B^{\rm halo}_0)^2}\right .\nonumber\\
  & &\left . + \frac{(B^{\rm halo}_x)^2+(B^{\rm halo}_y)^2+(B^{\rm halo}_z)^2}{(B^{\rm
      disc}_0)^2+(B^{\rm halo}_0)^2} \right ).
\end{eqnarray}
The Stokes $Q$ integration along the line-of-sight is
\begin{eqnarray}
  Q,U(y+\Delta y) & = & Q,U(y) + n_{\rm CR} \cdot \nonumber\\
& &  \left \lbrace \left ( (  B^{\rm disc}_x)^2 + (B^{\rm
    disc}_z)^2 \right ) \cdot \cos(2 \psi^{\rm Faraday}_{\rm disc})
\right . \nonumber\\
& & + \left . \left ( (  B^{\rm halo}_x)^2 + (B^{\rm
    halo}_z)^2 \right ) \cdot\ \cos(2 \psi^{\rm Faraday}_{\rm halo})
\right \rbrace ,
\end{eqnarray}
where for Stokes $U$, ``sin'' is used instead of ``cos''. Stokes $Q$
and $U$ have to be calculated at at least two frequencies in order to
calculate the rotation measure. The two-dimensional arrays of $Q(x,z)$ and
$U(x,z)$ are written out as a {\tt fits} file for which we used {\tt
  cfitsio}\footnote{CFITSIO is a library of C and Fortran sub-routines for
  reading and writing data files in FITS (Flexible Image Transport System)
  data format. CFITSIO is free software and can be obtained at {\tt
    http://heasarc.gsfc.nasa.gov/fitsio}}.

\end{document}